\documentclass[a4paper,superscriptaddress,twocolumn,showpacs,amsmath,amssymb,floatfix]{revtex4}

\usepackage{graphicx}
\usepackage{amssymb}
\usepackage{float}
\usepackage{hyperref}
\hypersetup{
	colorlinks=true,       
	linkcolor=blue,          
	citecolor=blue,        
	filecolor=blue,      
	urlcolor=blue           
}

\usepackage{color}

\input{epsf}

\begin{document}

\title{Thermoelectricity modeling with cold dipole atoms \\
in Aubry phase of optical lattice}

\author{Oleg V. Zhirov}
\affiliation{Budker Institute of Nuclear Physics,
630090 Novosibirsk, Russia}
\affiliation{Novosibirsk State University, 630090 Novosibirsk, Russia}
\affiliation{Novosibirsk State Technical University, Novosibirsk, Russia}
\author{Jos\'e Lages}
\affiliation{Institut UTINAM, OSU THETA, CNRS, 
Universit\'e Bourgogne Franche-Comt\'e, Besan\c con, France }
\author{Dima L. Shepelyansky}
\affiliation{\mbox{Laboratoire de Physique Th\'eorique, IRSAMC, 
Universit\'e de Toulouse, CNRS, UPS, 31062 Toulouse, France}}

\date{November 14, 2019}

\begin{abstract}
We study analytically and numerically the thermoelectric properties of
a chain of cold atoms with dipole-dipole interactions
placed in an optical periodic potential.
At small potential amplitudes the chain slides freely that 
corresponds to the Kolmogorov-Arnold-Moser phase of integrable curves
of a symplectic map.
Above a certain critical amplitude the chain is pinned by the lattice
being in the cantori Aubry phase. We show that the Aubry phase is 
characterized by exceptional thermoelectric properties
with the figure of merit $ZT = 25$ being ten times larger than the maximal
value reached in material science experiments.
We show that this system is well
accessible for magneto-dipole cold atom experiments
that opens new prospects for investigations of thermoelectricity.
\end{abstract}

%

\maketitle

\section{Introduction}
The phenomenon of Aubry transition 
describes the transport properties 
of a chain of particles linked by linear springs
in a periodic potential. At a small potential amplitude
the chain slides freely while above 
a certain potential amplitude
it is pinned by the potential.
This system is known
as the Frenkel-Kontorova model \cite{obraun}. 
The transition takes place at a fixed 
incommensurate density of particles per period
$\nu$. In fact the equilibrium positions
of particles are described 
by the Chirikov standard map \cite{chirikov,lichtenberg,meiss}
which represents a cornerstone model
of systems with dynamical chaos and symplectic maps.
Indeed, a variety of physical systems can be 
locally described by this map \cite{stmapscholar}.
In the frame of map description a density of particles
corresponds to a winding number of an
invariant Kolmogorov-Arnold-Moser (KAM) curve.
Such curves 
cover the main part of the phase space 
at small potential amplitudes
(small kick amplitudes $K$ of the map).
At large amplitudes the main part of phase space becomes chaotic and
the KAM curves are transformed
into cantori invariant sets which 
correspond to the chain ground states at given density
as it was proved by Aubry \cite{aubry}.
In addition to nontrivial mathematical properties,
the Frenkel-Kontorova model
represents a fundamental interest for
incommensurate crystals of interacting particles \cite{pokrovsky}.

An experimental realization of linear spring
interactions between particles
is not very realistic. Thus in \cite{fki2007}
it was proposed to consider a chain of Coulomb charges
placed in a periodic potential.
It was shown that this system of a one-dimensional (1D) Wigner crystal 
in a periodic potential can be locally described 
by the Chirikov standard map and the Frenkel-Kontorova model.
Thus the Aubry-like transition from the sliding
KAM phase to the Aubry pinned phase 
takes place at a certain critical potential amplitude
$K_c$. The dependence of $K_c$ on the density $\nu$  is
obtained by a local description in terms of
the Chirikov standard map \cite{fki2007,lagesepjd}.
For an experimental realization of the Aubry-like transition
it was proposed to use cold ions placed in both a periodic
potential and a global harmonic trap \cite{fki2007}.
The experimental studies of such a system 
had been started in \cite{haffner2011,vuletic2015sci}.
The first signatures of the Aubry transition have been reported 
by the  Vuletic group with up to
5 ions \cite{vuletic2016natmat}. 
The chains with a larger number of ions are now 
under investigations in \cite{ions2017natcom,drewsen}. 

A significant interest for a
Wigner crystal transport 
in a periodic potential
is related to the recent results showing that the Aubry
phase is characterized by remarkable thermoelectric properties with
high Seebeck coefficient $S$ and high figure of merit 
$ZT$ \cite{ztzs,lagesepjd}. 
The fundamental grounds of thermoelectricity had been 
established in far 1957 by Ioffe \cite{ioffe1,ioffe2}.
The thermoelectricity  is characterized by 
the Seebeck coefficient $S=-\Delta V /\Delta T$
(or thermopower). It is expressed through a voltage difference $\Delta V$
compensated by a temperature difference $\Delta T$.
Below we use units with a charge $e=1$ and the Boltzmann constant $k_B=1$ 
so that $S$ is dimensionless ($S=1$ corresponds to
$S \approx 88 \rm\mu V/K$ (microvolt per Kelvin)).
The thermoelectric materials are ranked by a figure of merit 
$ZT=S^2\sigma T/ \kappa$ \cite{ioffe1,ioffe2}
where $\sigma$ is the electric conductivity,
$T$ the temperature, and $\kappa$ the 
material thermal conductivity.

At present, the request for efficient energy usage stimulated extensive 
investigations of various materials with high
characteristics of thermoelectricity as reviewed in
\cite{sci2004,thermobook,baowenli,phystod,ztsci2017}.
The request is to find materials with $ZT >3$
that would allow an efficient conversion 
between electrical and thermal forms of energy.
The best thermoelectric materials created till now have $ZT \approx 2.6$.
At the same time, the numerical modeling reported 
for a Wigner crystal in the Aubry phase
reached values $ZT \approx 8$ \cite{ztzs,lagesepjd}. 

Thus investigations of Wigner crystal transport in a periodic potential
can help to understand the conditions favoring high $ZT$ values.
At present, a hundred of cold trapped-ions can be routinely kept for
hours in experimental device \cite{qcionrev} and thus such a system
is promising for experimental investigations of 
thermoelectricity \cite{lagesepjd}.
However, for a typical distance between charges being $\ell \sim 1\mu\rm m$ 
the Coulomb interactions are rather strong and  very high amplitudes 
of optical lattice potential 
$V_A \sim k_B \times 3\rm K$ (Kelvin)
are required  \cite{lagesepjd}.  This is hard to reach experimentally 
since typical optical potential amplitudes are 
$V_A \sim k_B \times  10^{-3}\rm K$ 
\cite{schaetz}. Thus to find a more suitable experimental
realization of Aubry transition we study here a chain of magneto-dipole atoms
placed in an optical periodic potential. The strength of interactions
between nearby  magneto-dipole atoms on a distance of $1\mu\rm m$  
is significantly smaller compared to Coulomb interactions, and thus a
significantly smaller amplitude of the optical potential
is required for the observation of the Aubry-like transition.
Indeed, the experimental investigations of quantum properties of 
cold magneto-dipole atoms allowed to observe a number of interesting
many-body effects (see e.g. \cite{pfau1,pfau2,pfau3}).

\section{Methods} The chain of atoms with magneto-dipole interactions
in 1D periodic potential is described by the Hamiltonian:
\begin{equation}
\begin{array}{cll}
H &=& {\displaystyle\sum_{i=1}^N} \left( \displaystyle\frac{{P_i}^2}{2} + V(x_i) \right) + U_I \; ,\\
V(x_i) &=& - K  \cos x_i \; ,\qquad \; U_I = \displaystyle\sum_{i > j} \displaystyle\frac{U_{dd}}{{\mid x_i - x_j \mid}^3}
\end{array}
\label{eq:ham1d}
\end{equation}
Here $x_i,P_i$ are conjugated coordinate and momentum of
atom $i$,  and $V(x)$  is an external 
periodic optical potential of amplitude $K$. 
The magneto-dipole-dipole interactions are given
by the $U_I\propto U_{dd}= \mu^2/(\ell/2\pi)^3$ term
with $\mu \approx 10 \mu_B$
(for ${^{164}}\rm Dy$ atoms) and $\mu_B$ is the Bohr magneton
(we assume all magnetic momenta to be polarized) \cite{pfau1}.
The Hamiltonian
is written in dimensionless units
where the lattice period is $\ell=2\pi$
and atom mass is $m=1$.
In the following we also take $U_{dd}=1$
so that $K$ in (\ref{eq:ham1d}) is the dimensionless amplitude of the periodic
potential
and thus the physical interaction
is $U_A = U_{dd} K$.
In these atomic-type units, 
the physical system parameters are measured in
units of   $r_a= \ell/2\pi$ for length, and of 
$\epsilon_a = \mu^2/{r_a}^3=U_{dd}$ for energy.
For $\ell = 1 \mu\rm m$ the  dimensionless temperature $T=1$ (or $k_B T$)
corresponds to the physical temperature
$T=  \epsilon_a/k_B  = 100  {\mu_B}^2/(k_B (\ell/2\pi)^3) 
\approx 25 \rm nK $  (nano-Kelvin) at $\mu=10 \mu_B$.

The thermoelectric properties of model (\ref{eq:ham1d}) 
are determined in the framework of the Langevin approach \cite{ztzs,lagesepjd}
with the equations of motion:
$\dot{P_i}= \dot{v}_i= -\partial H/\partial x_i +f_{dc} -\eta P_i+g \xi_i(t)$,
$ \dot{x_i} = P_i  = v_i$. Here $\eta$  describes phenomenologically the 
dissipative relaxation processes, and 
the amplitude of Langevin force $g$ is given 
by the fluctuation-dissipation theorem $g=\sqrt{2\eta T}$;
the random variables $\xi_i$ have normal distribution being $\delta-$correlated,
$v_i$ is atom velocity, $f_{dc}$ is a static force applied to atoms. 
As in \cite{ztzs,lagesepjd},
we use $\eta=0.02$, the results being not sensitive to this quantity.
The computations of $S$ are done as it is described in
\cite{ztzs,lagesepjd}. At fixed temperature $T$,
we apply a static force $f_{dc}$ which
creates an energy (voltage) drop 
$\Delta V = f_{dc} 2\pi L$ and
a gradient of atom density $\nu(x)$
along the chain with $L$ potential periods and $N$ atoms. 
Then, for $f_{dc}=0$ within
the Langevin equations, 
we impose a linear gradient of temperature
$\Delta T$  along the chain, and in the stabilized 
steady-state regime, we determine the charge density
gradient of $\nu(x)$ along the chain
(see e.g. Fig. 2 in \cite{ztzs}).
The data are obtained in the linear regime
of relatively small $f_{dc}$ and $\Delta T$ values.
Then, the Seebeck coefficient, $S=\Delta V/\Delta T$,
is computed using values of $\Delta V$ and $\Delta T$ 
for which the density gradient obtained 
from the application of a voltage $\Delta V$  
compensates the one obtained from 
the application of a gradient of temperature $\Delta T$.
We used the computation times up to $t=10^8$
to achieve the relaxation of the chain and to reach the
required statistical accuracy.

We assume that the cold atoms are in contact with an external environment which is able to play the role of thermostat, e.g. residual gas, etc. In order to compute Seebeck coefficient, we need a temperature gradient along the chain. In the Langevin equation, we can impose that the temperature $T$ is a function of the atom position $x$ along the chain, $T=T(x)=T_0 + g x $, where $T_0$ is the average temperature and $g=dT/dx$ is a small temperature gradient. In cold atom experiments, such a temperature gradient can be setup by multiple laser beams generating a zero average fluctuating force $f(x)$. The average of the square of the force, $f^2(x)$, should change linearly along the chain. Consequently, these laser beams induced fluctuating forces will create additional ion velocity fluctuations with $(\delta v_i)^2 \propto  f^2(x) \propto (T(x)-T_0) = g x$ producing a temperature gradient along the ion chain.

\begin{figure}[t!]
	\centering
	\includegraphics[width=\columnwidth]{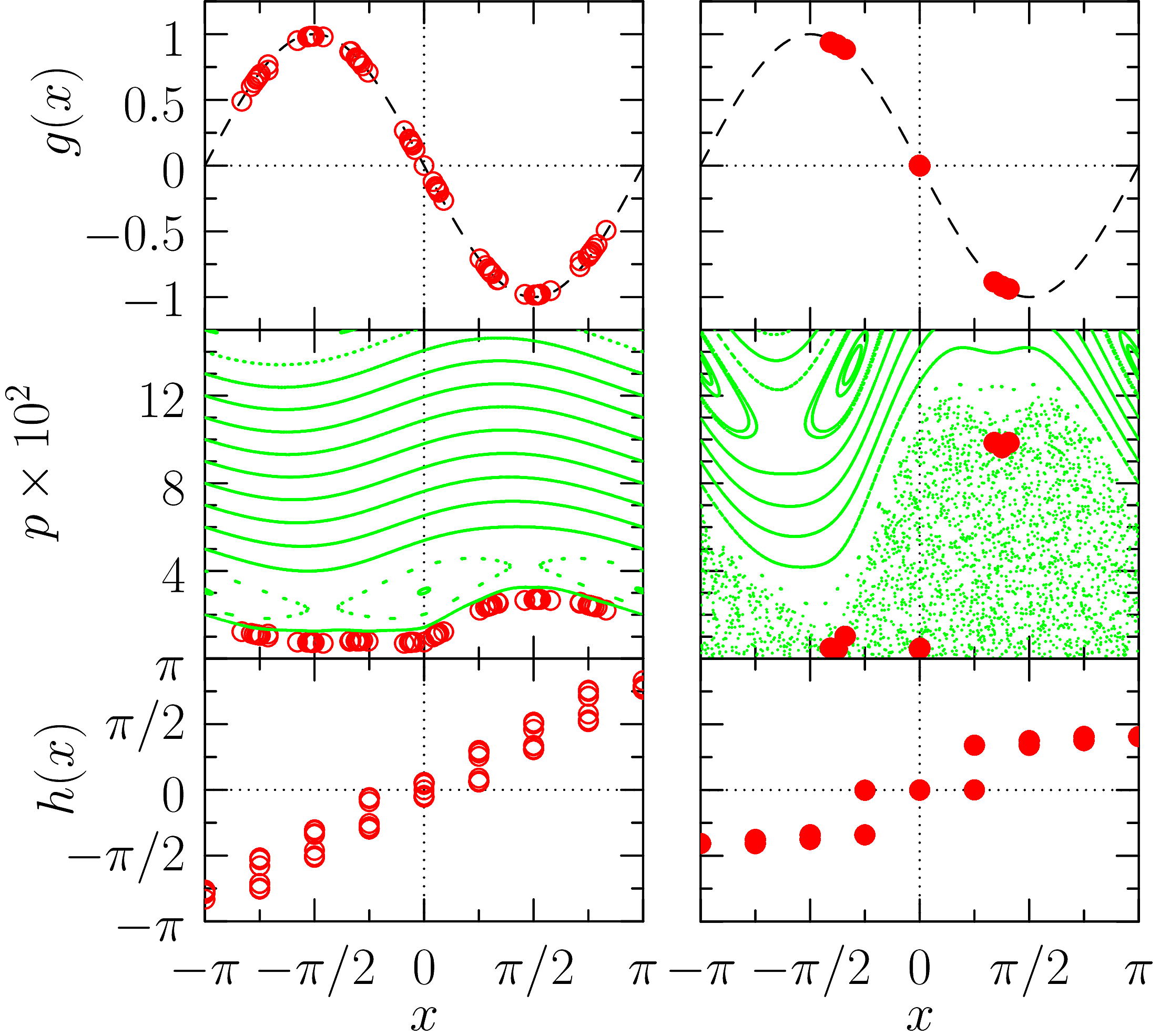}
	\caption{\label{fig1}
		Functions related to the dynamical map obtained from the ground state equilibrium positions
		$\left\{x_1,\dots,x_N\right\}$ of the atoms at $K = 0.02$ (left column) and $K = 0.1$ (right column). 
		Top row: the red circles show the reconstructed kick function $g(x)$ using the the ground state positions $\left\{x_1,\dots,x_N\right\}$. The dashed curved gives the theoretical form of the kick function $g(x)=-\sin x$.
		Middle row: phase space portrait of the $(x,p)$-map (\ref{eq:map}) (green points) and the
		actual ground state positions of the atoms (red circles).
		Bottom row: the atom positions  are shown via
		the hull function
		$h(x) = (x_i+\pi) [{\rm mod }\; 2\pi]-\pi$ versus
		$x=(2\pi (i-1)/\nu {+\pi} ) [{\rm mod }\; 2\pi] -\pi$; the positions of the two
		atoms
		at the chain ends are fixed.
		Here we have $N=89$ atoms and $L=55$ periods,
		$\nu=89/55$.
	}
\end{figure}

\section{Results}

\subsection{Ground state properties} The equilibrium static atom positions
are determined by the conditions $\partial H/ \partial x_i =0$, $P_i=0$ 
\cite{aubry,fki2007,lagesepjd}. In the approximation of nearest  neighbor
interacting atom, this leads to the symplectic map for recurrent
atom positions $x_i$
\begin{equation}
p_{i+1} = p_i + (K/3) g(x_i) \; , \; \; x_{i+1} = x_i+1/p_{i+1}^{1/4} \; ,
\label{eq:map}
\end{equation}
where the effective momentum $p_i = 1/(x_{i}-x_{i-1})^4$ is conjugated to $x_i$ 
and the kick function $g(x)$ is such as
$K g(x_i)/3= \left.-dV/dx\right|_{x=x_i}/3 = -(K/3) \sin x_i$.
This map description assumes only nearest neighbor interactions.
Below, we show that it well describes the real situation 
with interactions between all the atoms.
This is rather natural since the nearest neighbor approximation worked already well
for ions with Coulomb interactions \cite{fki2007,lagesepjd}, moreover for dipole atoms, the interactions drop even faster
with the distance between atoms.

As in \cite{ztzs,lagesepjd}, the validity of the map (\ref{eq:map})
is  checked numerically by finding 
the ground state configuration using numerical methods
of energy minimization 
described in \cite{aubry,fki2007} and taking 
into account the long range nature of the interactions between all the atoms. The obtained ground state positions, $\left\{x_1,\dots,x_N\right\}$, of the $N$ atoms allows to recursively determine the $N$ effective momenta, $\left\{p_1,\dots,p_N\right\}$, by inverting the second equation of the map (\ref{eq:map}), $p_i=\left(x_{i+1}-x_i\right)^{-4}$. Once obtained, the effective momenta can be used to compute successively $N$ values $\left\{g(x_1),\dots,g(x_N)\right\}$ of the kick function $g(x)$, by using the first equation of the map (\ref{eq:map}), $\left(K/3\right)g(x_i)=p_i-p_{i+1}$. Fig.~\ref{fig1} (top panels) shows that the map (\ref{eq:map}), obtained in the nearest neighbor approximation, is indeed valid since the $\left\{g(x_1),\dots,g(x_N)\right\}$ values fall on the top of the function $g(x)=\sin x$ analytically obtained from the equations of motion derived from the Hamiltonian (\ref{eq:ham1d}). This means that the description of the positions of the $N$ atoms in the ground state can be mapped on a dynamical problem of a fictitious kicked particle taking successively the positions, $\left\{x_1,\dots,x_N\right\}$, and the momenta , $\left\{p_1,\dots,p_N\right\}$. The dynamics of the fictitious kicked particle is governed by the map (\ref{eq:map}). Phase space portraits of the map (\ref{eq:map}) is given in Fig.~\ref{fig1} (middle panels). A phase space portrait (green points and curves) is obtained by taking many different initial conditions $(x_0,p_0)$ and computing many of the corresponding successive phase space points $(x_i,p_i)$. In Fig.~\ref{fig1} (middle panels), we show the phase space portrait for KAM phase ($K=0.02<K_c$, left panel) and for the Aubry phase ($K=0.1>K_c$, right panel). We observe that the ground state positions of the atoms (red circles) are located, for $K=0.02$ (left panel), on the top of an invariant KAM curve which is a situation corresponding to a regular motion of the fictitious kicked particle, and, for $K=0.1$ (right panel), in the chaotic component of the phase space portrait which is a situation corresponding to a chaotic motion of the fictitious kicked particle. All the chaotic dynamics concepts used in the article are well documented in the literature (see e.g. \cite{lichtenberg,meiss}). Fig.~\ref{fig1} (bottom panels) shows the hull function  $h(x) = (x_i+\pi) [{\rm mod }\; 2\pi]-\pi$
with $x=(2\pi (i-1)/\nu +\pi ) [{\rm mod }\; 2\pi] -\pi$.
Indeed, for $K \rightarrow 0$, i.e. for a vanishing optical potential,
we have  $h(x)= x$. For $K<K_c$, the hull function $h(x)$ still follows this linear law with smooth deviations,
while, for $K>K_c$, we obtain the devil's staircase corresponding to the fractal cantori structure of the chaotic phase space. The transition from the smooth hull function, which is typical for a sliding chain in the KAM phase,
to the devil's staircase, which is typical for a pinned chain in the Aubry phase, is visible in Fig.~\ref{fig1} (bottom panels). Thus, with $\nu= 89/55\simeq\nu_g= (1+\sqrt{5})/2=1.618...$, i.e. 89 atoms (in their ground state) distributed in 55 optical potential wells, the Aubry transition takes place at a
certain $K_{c}$ inside the interval $0.02 < K_{c} < 0.1$.

\begin{figure}[!t]
	\centering
	\includegraphics[width=0.49\columnwidth]{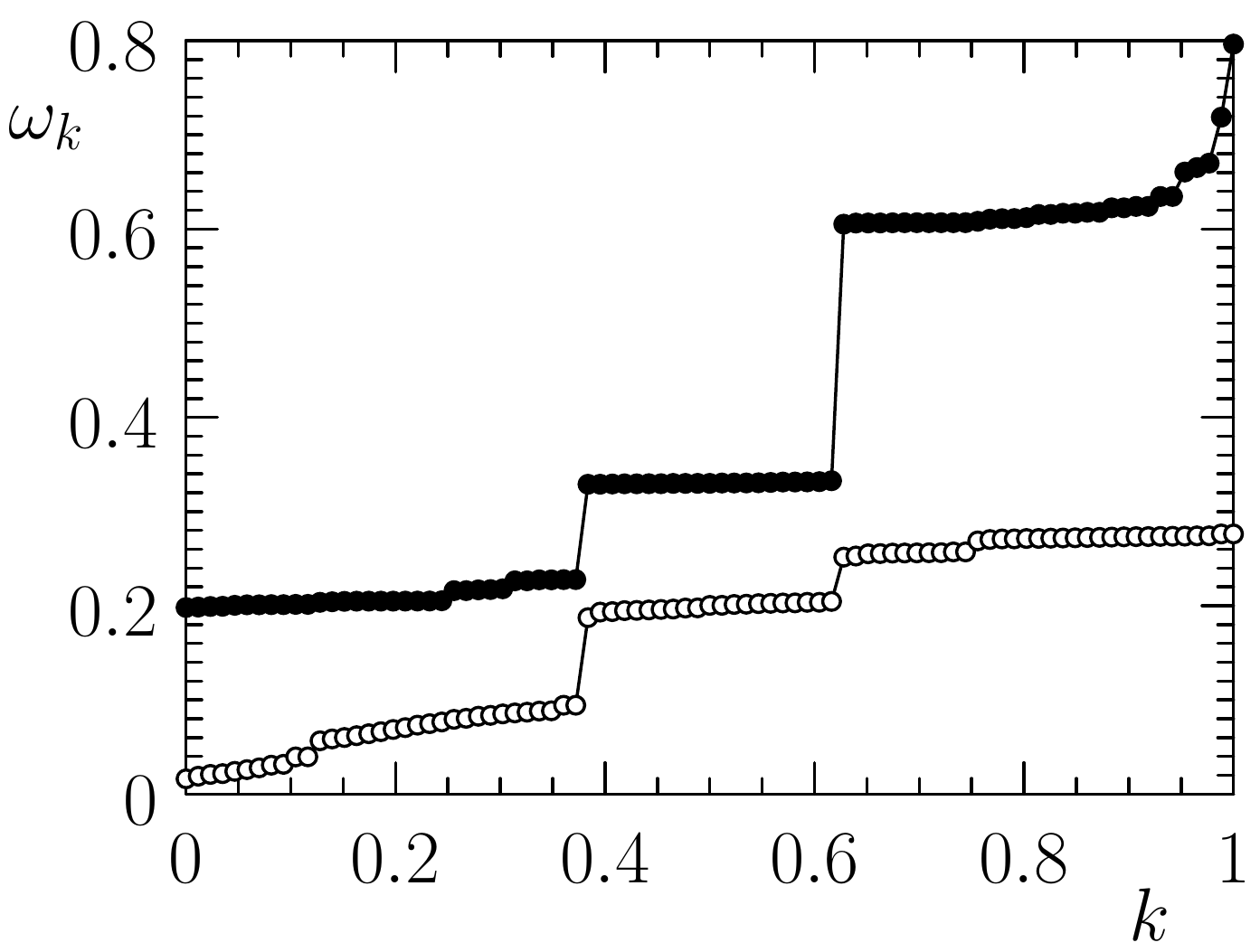}
	\includegraphics[width=0.49\columnwidth]{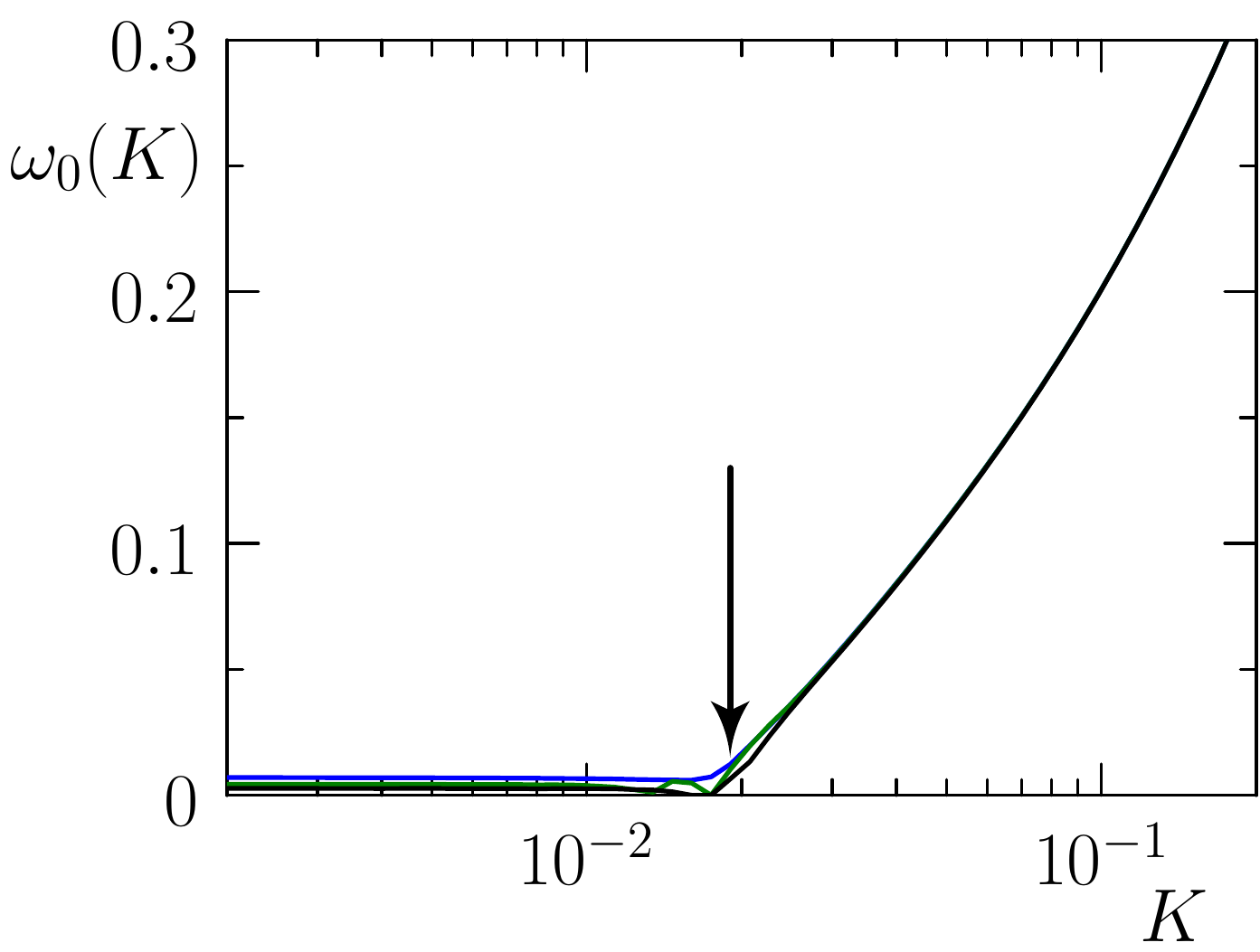}
	\caption{\label{fig2}Left panel: phonon spectrum $\omega(k)$ for $K=0.02$ (open circles) and $K=0.1$ (full circles) for $\nu=N/L=89/55$. Right panel: dependence of the lowest phonon frequency
		$\omega_0$ on $K$ for the lattice with
		$N$ dipole atoms in $L$ lattice periods at density
		$\nu \approx\nu_g= 1.618...$:
		$N/L = 55/34$ (blue curve), $89/55$ (green curve),
		$144/89$ (black curve); the arrow marks the point of the Aubry transition
		at $K_c=0.019$ defined as it is described in  the text. }
\end{figure}

\begin{figure}[t!]
	\centering
	\includegraphics[width=\columnwidth]{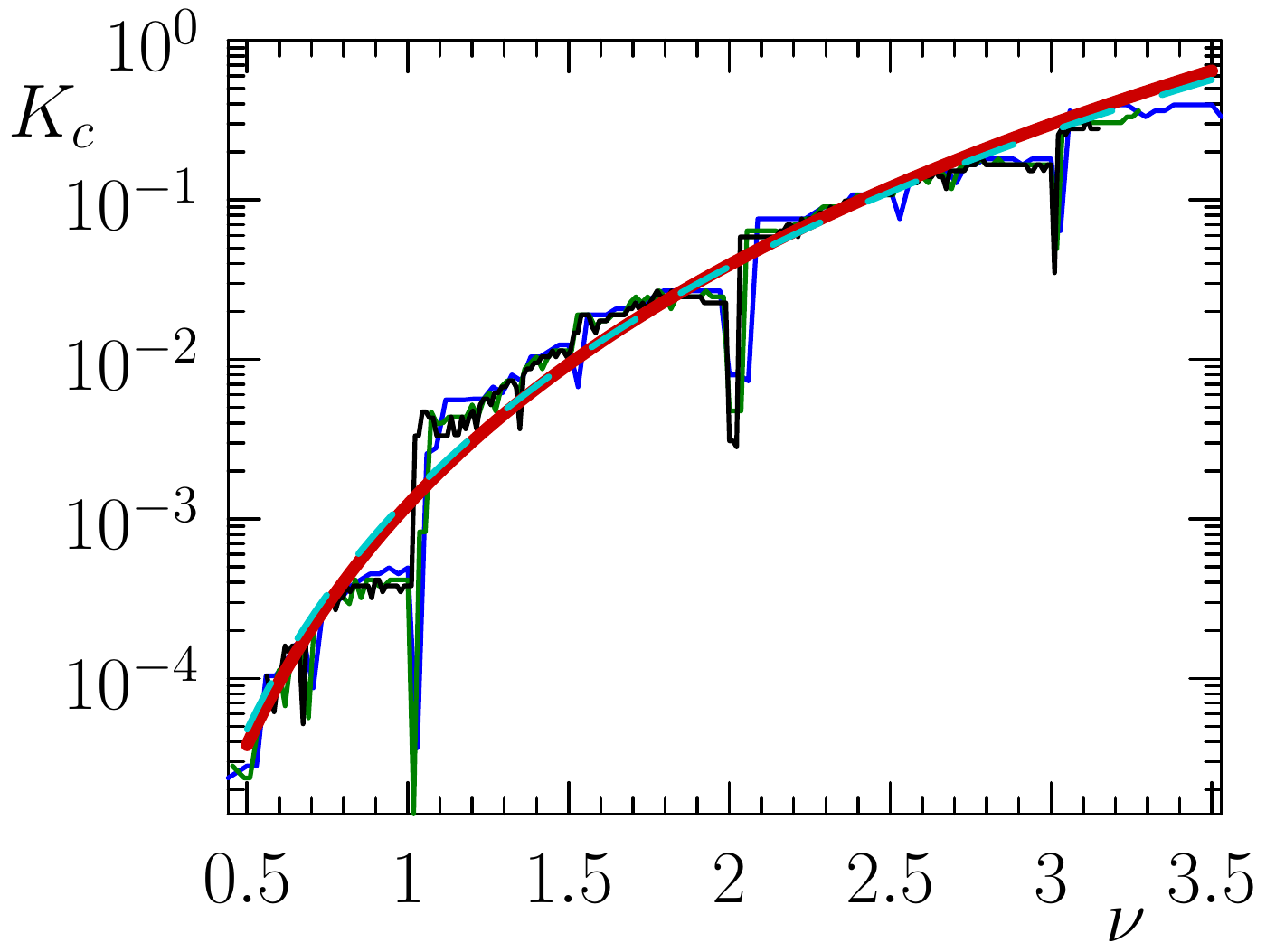}
	\caption{\label{fig3}
		Dependence of the Aubry transition critical threshold $K_c$ on the
		atomic density $\nu$ shown for different 
		chain lengths $L=34$ (blue curve), $55$ (green curve)
		and $89$ (black curve). The red full curve
		shows the theoretical dependence (\ref{eq:kc}),
		the  dashed cyan curve shows the fit
		with $K_c=0.0137 (\nu/\nu_g)^\alpha$, $\nu_g=1.618...$
		and $\alpha = 4.82 \pm 0.08$.
	}
\end{figure}

The equations of motion can be linearized in a vicinity
of equilibrium positions and, in this way, we obtain the phonon spectrum $\omega(k)$
of small oscillations with $k=i/N$ being a scaled mode number. 
The examples of spectrum are shown in Fig.~\ref{fig2} (left panel).
At $K=0.02$, in the KAM phase,
we have $\omega \propto k$ corresponding to the acoustic modes,
while at $K=0.1$, inside the Aubry phase, we have appearance of an optical spectral gap
related to the atomic chain pinned by the potential.
Such a modification of the spectrum properties is similar to the cases of the
Frenkel-Kontorova model \cite{aubry} and the ion chain \cite{fki2007,lagesepjd}. 
Fig.~\ref{fig2} (right panel) shows that the minimal spectral frequency $\omega_0(K)$ 
is practically independent of the optical potential amplitude $K$  inside the KAM phase at $K<K_c$ (being close to zero
with $\omega_0 \propto 1/L$) and it increases with $K$ inside the Aubry phase
being independent of $L$ for $K>K_c$. Thus, the critical $K_c$ values can be approximately determined
as an intersection of a horizontal line $\omega_0 = const$
with a curve of growing $\omega_0(K)$ at $K>K_c$.
From these properties, we obtain numerically that $K_c \approx 0.019$ for 
the Fibonacci density $\nu \approx \nu_g= 1.618...$ We obtained also $K_c \approx 0.14$ for $\nu = 2.618$,
and $K_c \approx 0.4$ for  $\nu = 3.618$.

\subsection{Density dependence of Aubry transition}
The dependence of Aubry transition point $K_c(\nu)$
can be obtained from the local description of the map 
(\ref{eq:map}) by the Chirikov standard map.
For that, the equation of $x_{i+1}$ in (\ref{eq:map})
is linearized in $p_{i+1}$ near the resonant value $p_r$ that leads to the
standard map form
$y_{i+1}=y_i - K_{\rm eff} \sin x_i$, $x_{i+1}=x_i-y_{i+1}$
with $y_i \propto p_i$ and $K_{\rm eff} =  (2\pi)^5 K /(12 \nu^5)$
(see details in Appendix \ref{appendixB} and \cite{lagesepjd}).
The last KAM curve is destroyed at $K_{\rm eff} \approx 1$ 
\cite{chirikov,lichtenberg} that gives
\begin{equation}
K_c \approx 12 (\nu/2\pi)^5 \approx 0.0136 (\nu/\nu_g)^5 \;, \; \nu_g=1.618... \;
\label{eq:kc}
\end{equation}

The numerically obtained dependence $K_c(\nu)$ is shown in Fig.~\ref{fig3}
for 3 different chain lengths $L$. On average, it is well described by the theory (\ref{eq:kc})
taking into account that $K_c$ is changed by almost 4 orders of magnitude
for the considered range $0.5 \leq \nu \leq 3.5$. There are certain deviations in the vicinity
of integer resonant densities $\nu=1,2,3$ which should be attributed to the presence of
a chaotic separatrix layer at such resonances that reduces significantly the critical
$K_c$ for KAM curve destruction (similar effect has been discussed for the ion chains \cite{lagesepjd}).

\begin{figure}[t]
	\centering
	\includegraphics[width=0.49\columnwidth]{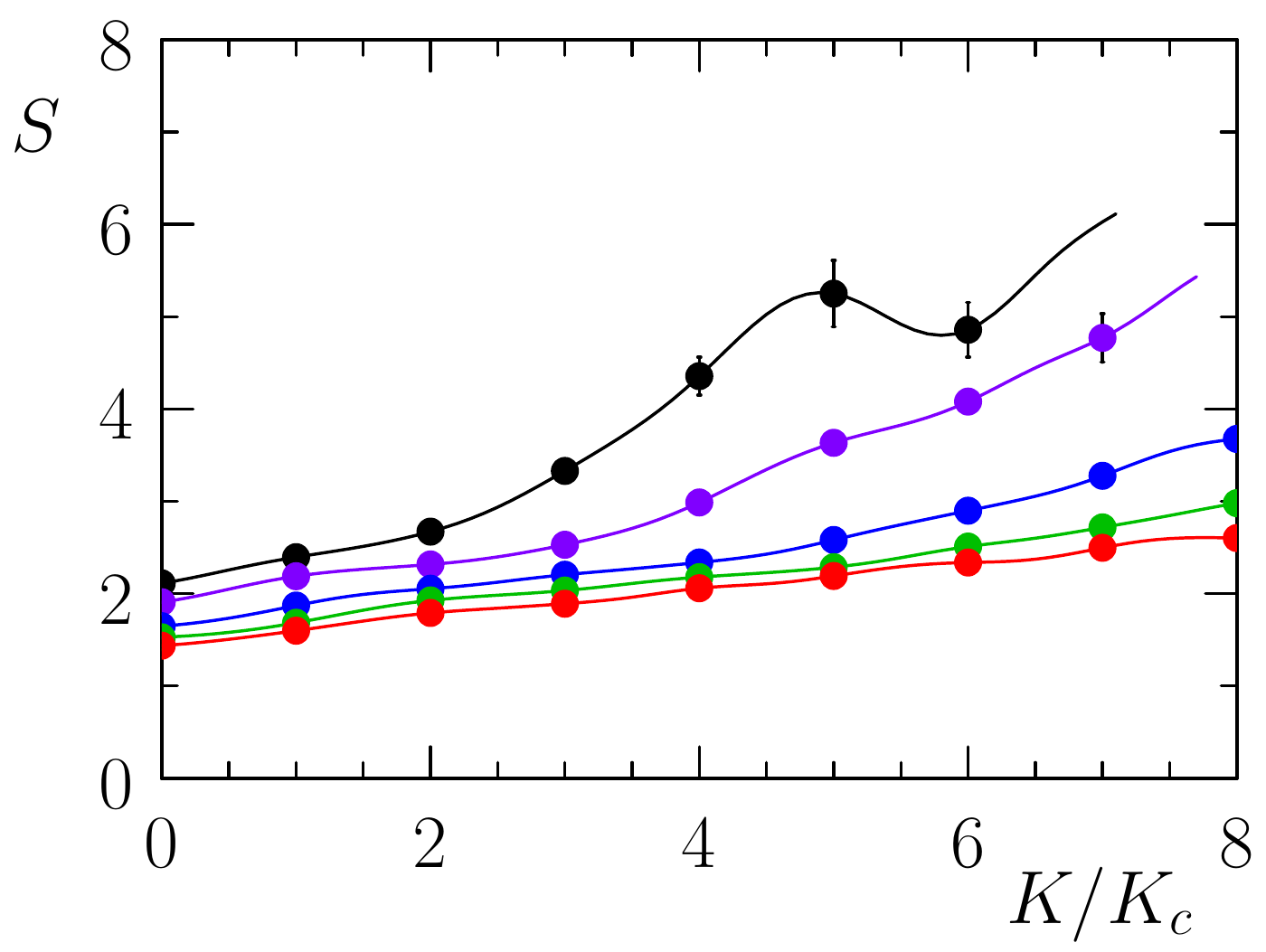}
	\includegraphics[width=0.49\columnwidth]{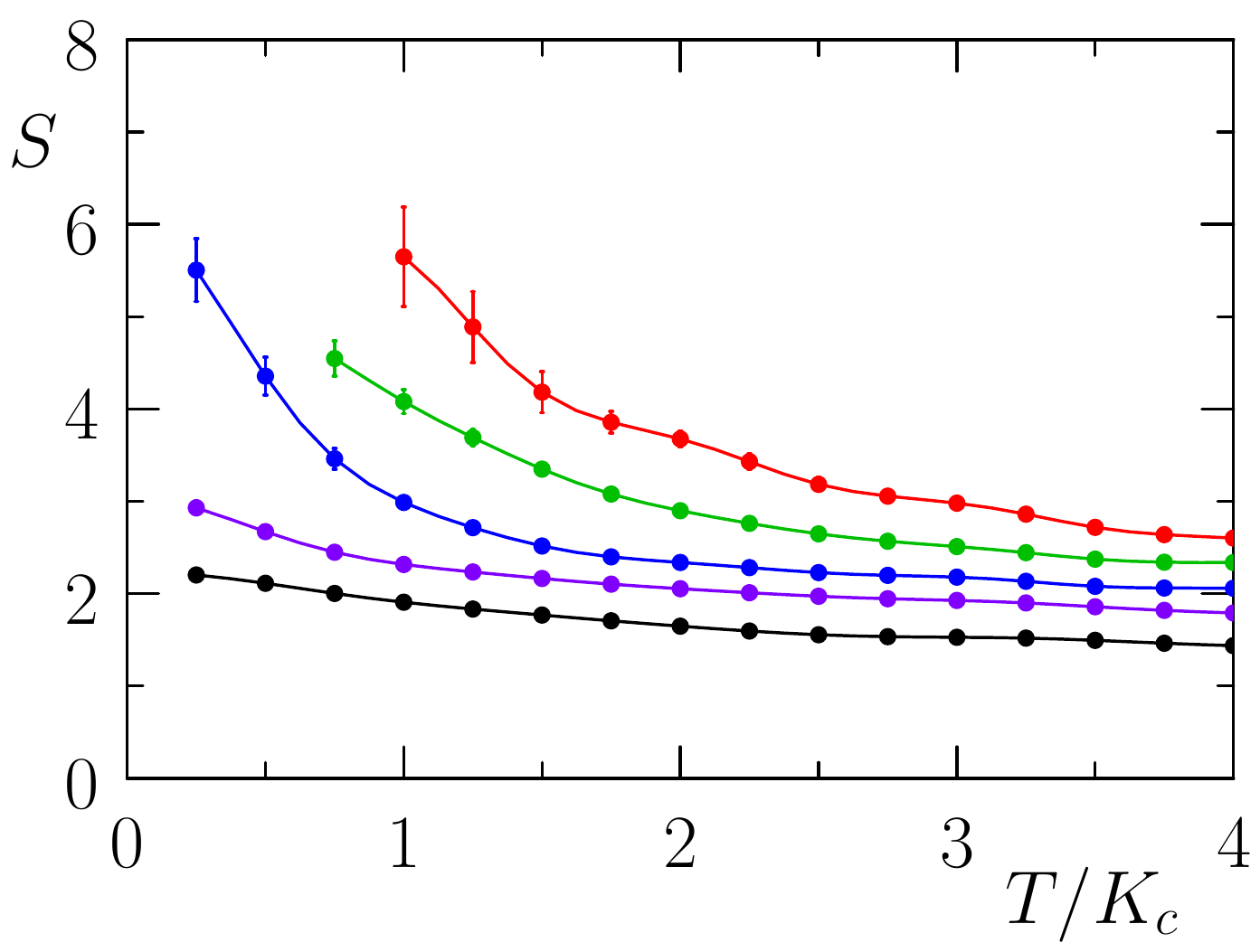}\\
	\includegraphics[width=0.49\columnwidth]{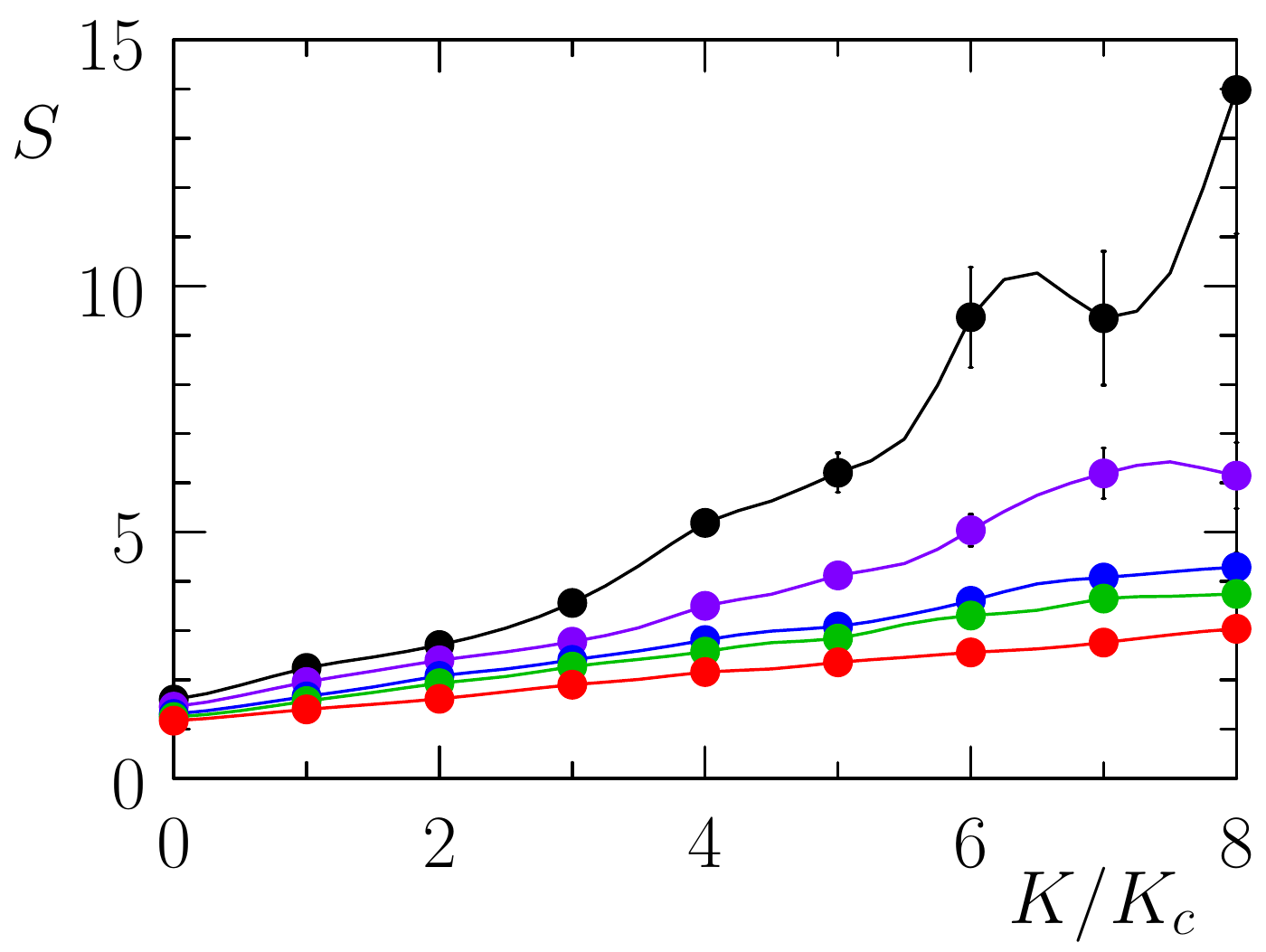}
	\includegraphics[width=0.49\columnwidth]{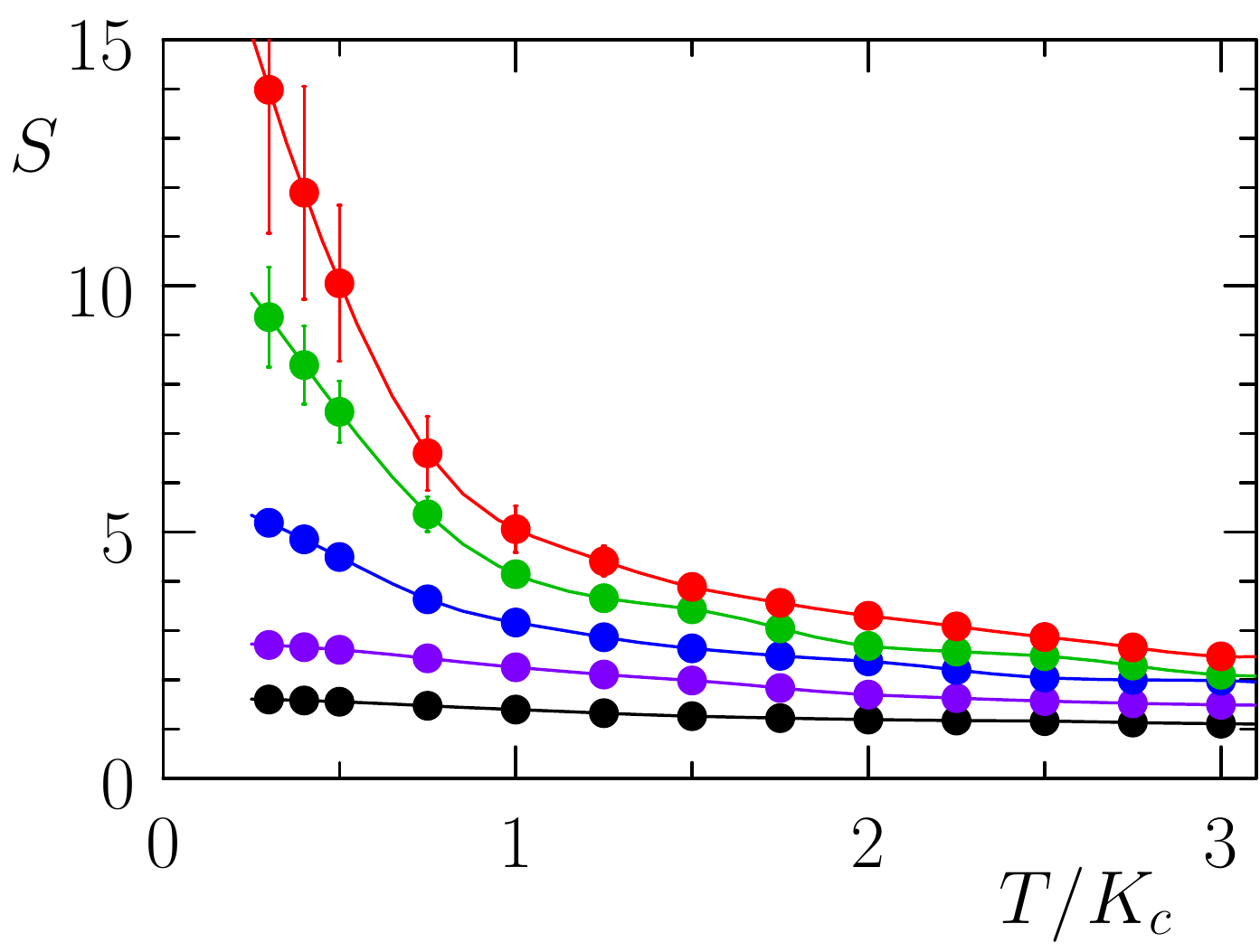}
	\caption{\label{fig4}
		Dependence of the Seebeck coefficient $S$ on the rescaled
		potential amplitude $K/K_c$ (left column)
		and temperature $T/K_c$ (right column)
		for $N/L=34/21 = \nu \approx 1.618$ (top row)
		and $N/L=55/21 = \nu \approx 2.618$ (bottom row).
		Left: color curves show the cases $T/K_c = 0.5, 1, 2, 3, 4$
		for $N/L=34/21$
		and $T/K_c = 0.3, 0.8, 1.3, 1.6, 2.3$
		for $N/L=55/21$
		(black, violet, blue, green, red from top to bottom curves).
		Right: $K/K_c= 0, 2, 4, 6, 8$ for $N/L=34/21$
		and  $N/L=55/21$
		(black, violet, blue, green, red from bottom to top curves).
		Here, $K_c = 0.019$ for $\nu=34/21$ and $K_c=0.14$ for $\nu = 55/21$.
	}
\end{figure}

\subsection{Seebeck coefficient} The dependencies of $S$ on potential amplitude
and temperature are shown in Fig.~\ref{fig4}
for two Fibonacci-like values of density $\nu \approx 1.618$ and $\nu \approx 2.618$.
The results clearly show that in the KAM phase
$K<K_c$ we have only rather moderate values of $S \sim 1$
being close to those value of noninteracting particles
(similar result was obtained in \cite{ztzs,lagesepjd}). 
In the Aubry phase at $K>K_c$, we have an increase of $S$ with the increase of $K/K_c$,
and a decrease of $S$ with the increase of $T/K_c$. This is rather natural since
at $T \gg K_c$ the lattice pinning effect disappears due 
to the fact that the atom energies become significantly larger than the barrier height,
and we approach to the case without potential corresponding
to the KAM phase. The maximal obtained values are as
high as $S \approx 10 - 15$ being still smaller those
obtained for ion chains \cite{ztzs,lagesepjd}.
Nevertheless, as shown below, we find very large figure of merit $ZT$ values
in this regime.

\begin{figure}[t!]
	\centering
	\includegraphics[width=\columnwidth]{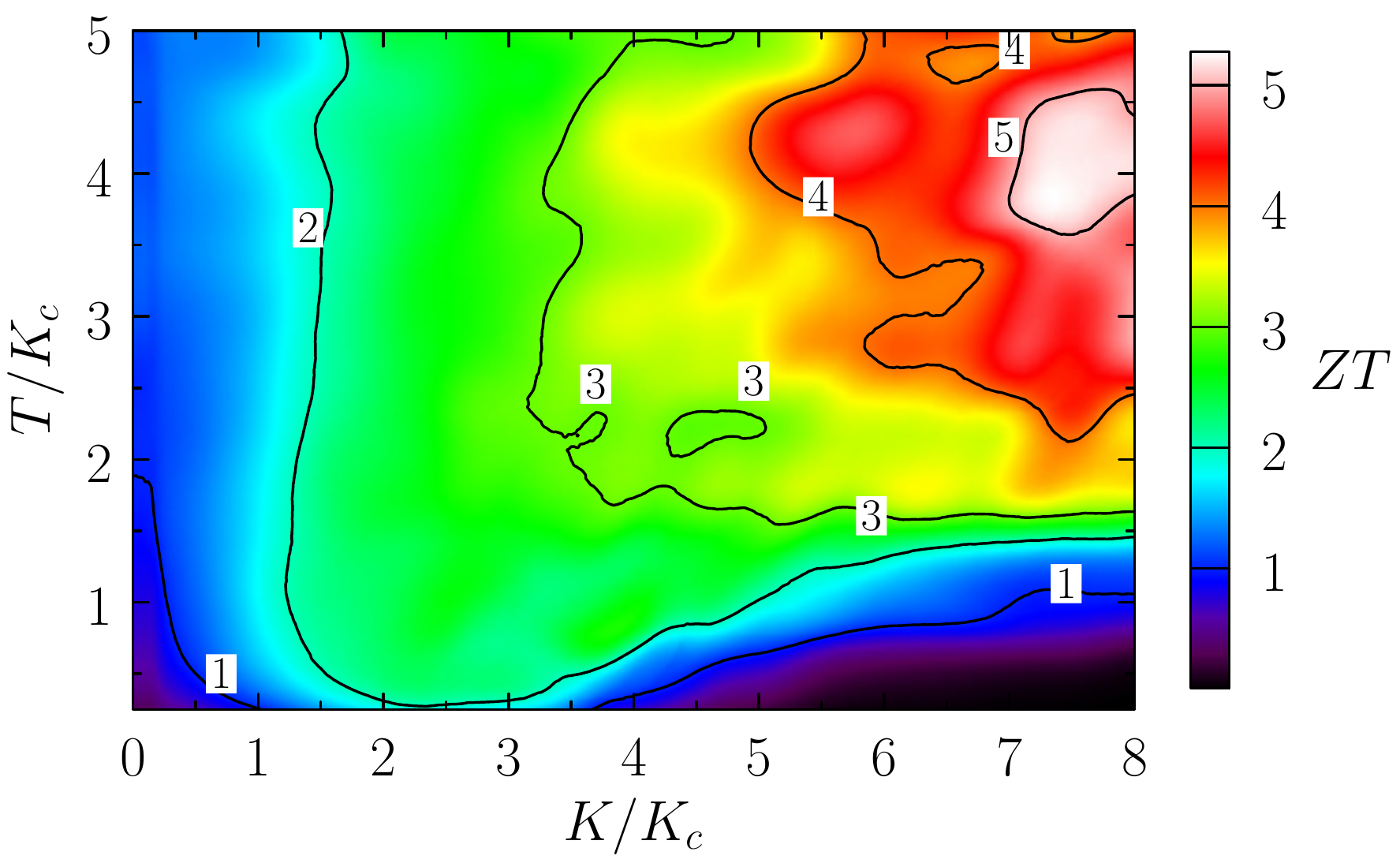}\\
	\includegraphics[width=\columnwidth]{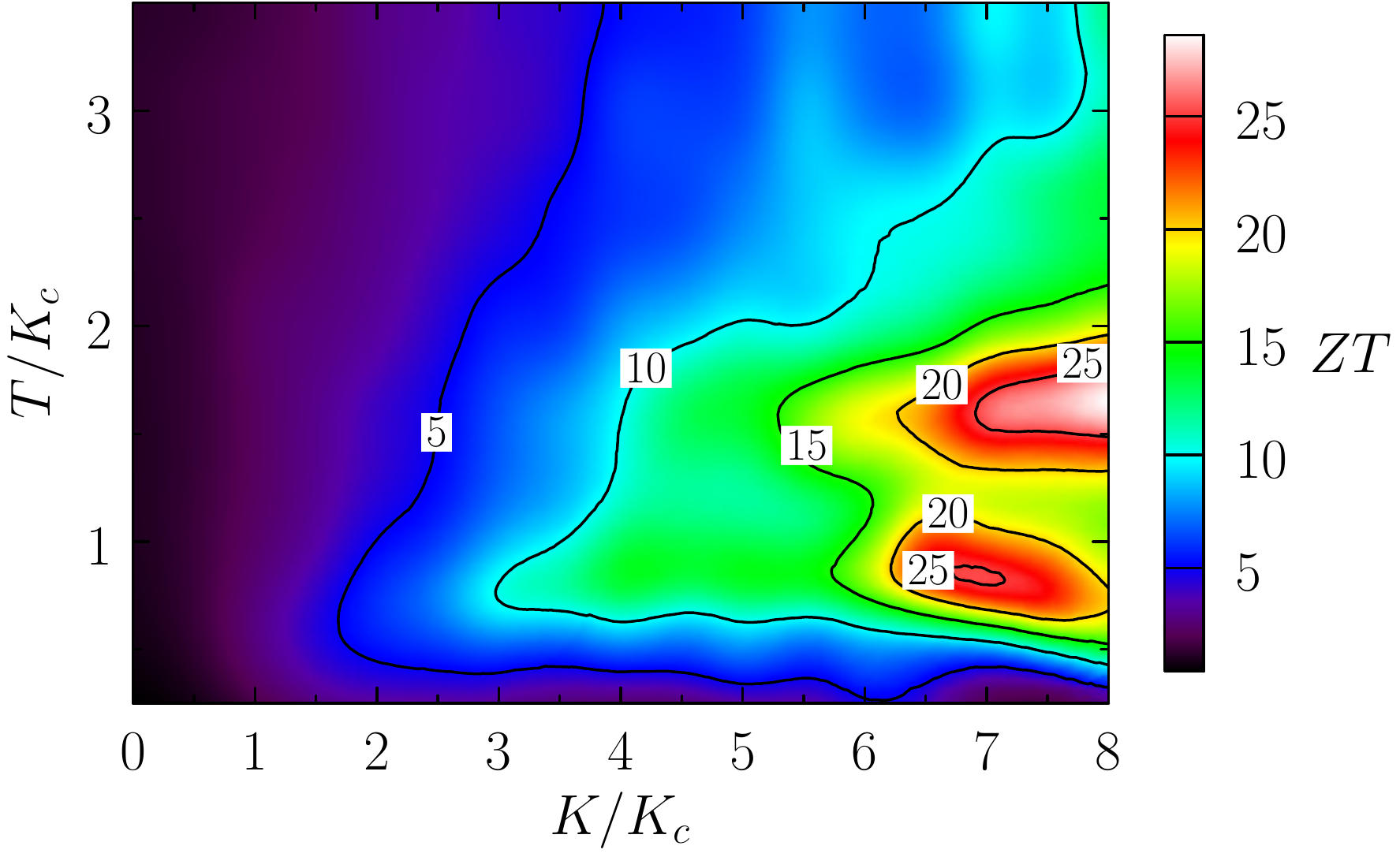}
	\caption{\label{fig5}
		Color map of the figure of merit $ZT$ values 
		for density $\nu= N/L= 34/21 \approx 1.618$ (top panel) and
		$\nu= N/L= 55/21 \approx 2.618$ (bottom panel);
		numbers mark isocontour values of $ZT$.
		Here, $K_c = 0.019$ for $\nu=34/21$ and $K_c=0.14$ for $\nu = 55/21$.
	}
\end{figure}

\subsection{Figure of merit}

To obtain the value of 
$ZT$ we need to compute the conductivity of atoms,
$\sigma$, and their thermal conductivity, $\kappa$.
The value of $\sigma$ is defined through
the current $J$ of atoms induced by 
a static force $f_{dc}$ for a chain with 
periodic boundary conditions: $j=\nu v_{at}/2\pi$
where $v_{at}$ is the average velocity of atoms
and $\sigma=j/f_{dc}$; for $K=0$, we have
$\sigma=\sigma_0 = \nu/(2\pi \eta)$.
The heat flow $J$ is induced by the temperature
gradient due to the Fourier law
with the thermal conductivity $\kappa = J/(\partial T /\partial x)$.
The heat flow is computed as it is described in
\cite{ztzs,lagesepjd} and in Appendix \ref{appendixC}.
The values of $\sigma$ and $\kappa$ drop
exponentially with the increase of $K$ inside the Aubry phase
at $K>K_c$ (see Figs.~\ref{figS2} and \ref{figS3} in Appendix \ref{appendixC}).
The computations performed for various
chain lengths at fixed density confirm that
the obtained values of $S, \sigma, \kappa$
are independent of the chain length for $T > K_c$
confirming that the results are obtained
in the thermodynamic limit
(see Fig.~\ref{figS4} in Appendix \ref{appendixD}).
For $T \ll K_c$, the pinning is too strong and 
much larger computation times are needed for numerical simulations
to reduce the fluctuations.

We note that, for cold atoms in an optical
lattice, a static force can be created by 
a modification of the lattice potential or by lattice acceleration.
There is now a significant progress with the temperature control of cold ions
and atoms (see e.g. \cite{haffner2014,paz,grimm}) 
and we expect that a generation of temperature gradients for measurements 
of $\kappa$ and $S$ can be realized experimentally.

With the computed values of $\sigma, \kappa, S$ ,
we determine the figure of merit $ZT$. Its dependence on
$K$ and $T$ is shown in Fig.~\ref{fig5} for 
$\nu \approx 1.618$ and $2.618$ (additional data are given in 
Figs.~\ref{figS5} and \ref{figS6} in Appendices \ref{appendixE} and \ref{appendixF}).
For $\nu \approx 1.618$, we obtain the maximal values
$ZT \approx 5$ being comparable with those of ion chain
reported in \cite{ztzs}. However, for $\nu \approx 2.618$,
we find significantly larger maximal values with $ZT \approx  25$.
We attribute such large $ZT$ values to the significantly more rapid spatial drop
of dipole interactions comparing to the Coulomb case, 
arguing that this produces a rapid decay of the heat conductivity with the
increase of $K > K_c$.

\section{Discussion} The obtained results show
that a chain of dipole atoms in a periodic potential
is characterized by outstanding values of the figure
of merit $ZT \approx 25$ being by a factor ten larger than the
actual $ZT$ values reached till present in material science \cite{ztsci2017}.
Thus, the experiments with cold dipole atoms in the Aubry phase of an
optical lattice open new prospects for
experimental investigation of fundamental aspects of thermoelectricity.

We note that for a laser wavelength $\lambda = 564\rm nm$, 
the optical lattice period is $\ell =\lambda/2 = 282 \rm nm$,
and thus for $\nu \approx 2.6$, we have the Aubry transition at the
potential amplitude $V_A/k_B =T_A= 0.14 \epsilon_a/k_B \approx   200 \rm nK$.
Such potential amplitude and temperature are 
well reachable with experimental setups at $T \approx 20\rm nK$ used
in \cite{pfau3}. 
At the same time at $T \approx   200 \rm nK$ the wave length of Dy atoms
becomes $\lambda_{\rm Dy} =\hbar/\sqrt{2m_{\rm Dy}k_BT} \approx 90  \rm nm < \ell$ 
being only a few times smaller
than the lattice period $\ell$. Thus the quantum effects can start to play
a role. But their investigations require a separate study.

Since such Aubry phase parameters
are well accessible for experiments, it may be also
interesting to test the quantum gate operations
of atomic qubits, formed by two atomic levels,
with dipole interaction between qubits.
Such a system is similar to ion quantum computer
in the Aubry phase discussed in \cite{qcion}.
As argued in \cite{qcion,loye20}, the optical gap
of Aubry phase should protect gate accuracy.

\section{Acknowledgments}
We thank D.Guery-Odelin for useful discussions about dipole gases.
This work was supported in part by the Pogramme Investissements
d'Avenir ANR-11-IDEX-0002-02, 
reference ANR-10-LABX-0037-NEXT (project THETRACOM).
This work was also supported in part by the Programme Investissements
d'Avenir ANR-15-IDEX-0003, ISITE-BFC (project GNETWORKS), 
and in part by the Bourgogne Franche-Comt\'e Region
2017-2020 APEX Project.



\appendix

\section{Local description by the Chirikov standard map }
\label{appendixB}

The local description of the map (\ref{eq:map}) 
by the Chirikov standard map is done in the standard way:
the second equation for the phase change of $x_{i+1}$
is linearized in the momentum $p_{i+1}$ near the 
resonances defined by the condition
$x_{i+m}-x_i = mp_r^{-1/4}=2\pi m'$ where $m$ and $m'$ are integers.
This determines the resonance positions in momentum
with $p_r(\nu) =  (\nu/2\pi)^4$ with $\nu=m/m'$.
Then the Chirikov standard map
is obtained with the variables
$y_i= \alpha_r (p_i - p_r)+p_r^{-1/4}$, $\alpha_r=1/(4{p_r}^{5/4})$
and $K_{\rm eff}=K(2\pi/\nu)^5/12$. The critical $K_{\rm eff}$ value
is defined by the condition $K_{\rm eff} \approx 1$
that leads to the equation (\ref{eq:kc}).

\section{Numerical computation of conductivity}
\label{appendixC}

\begin{figure}[!h]
	\begin{center}
		\includegraphics[width=0.49\columnwidth]{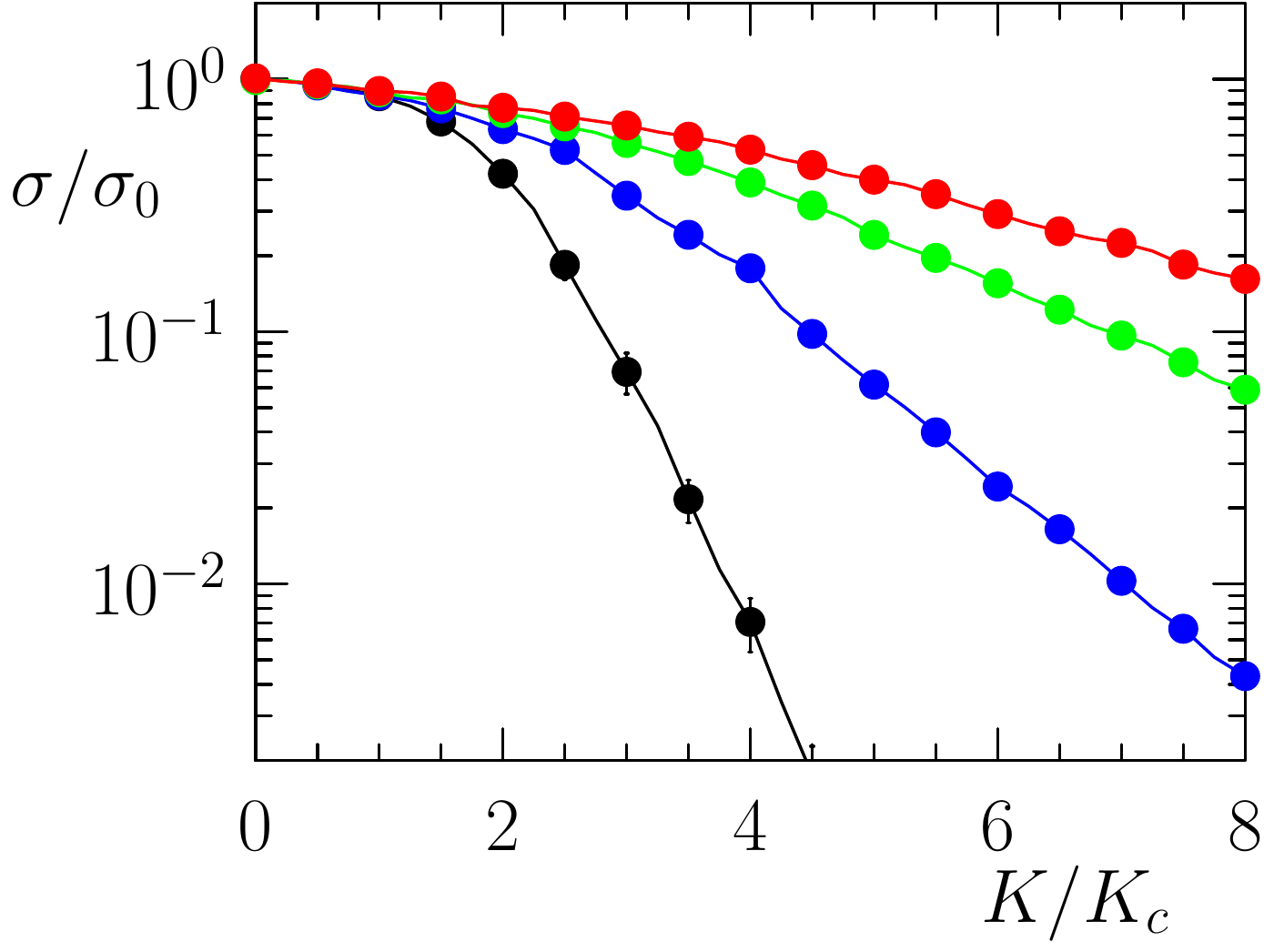}
		\includegraphics[width=0.49\columnwidth]{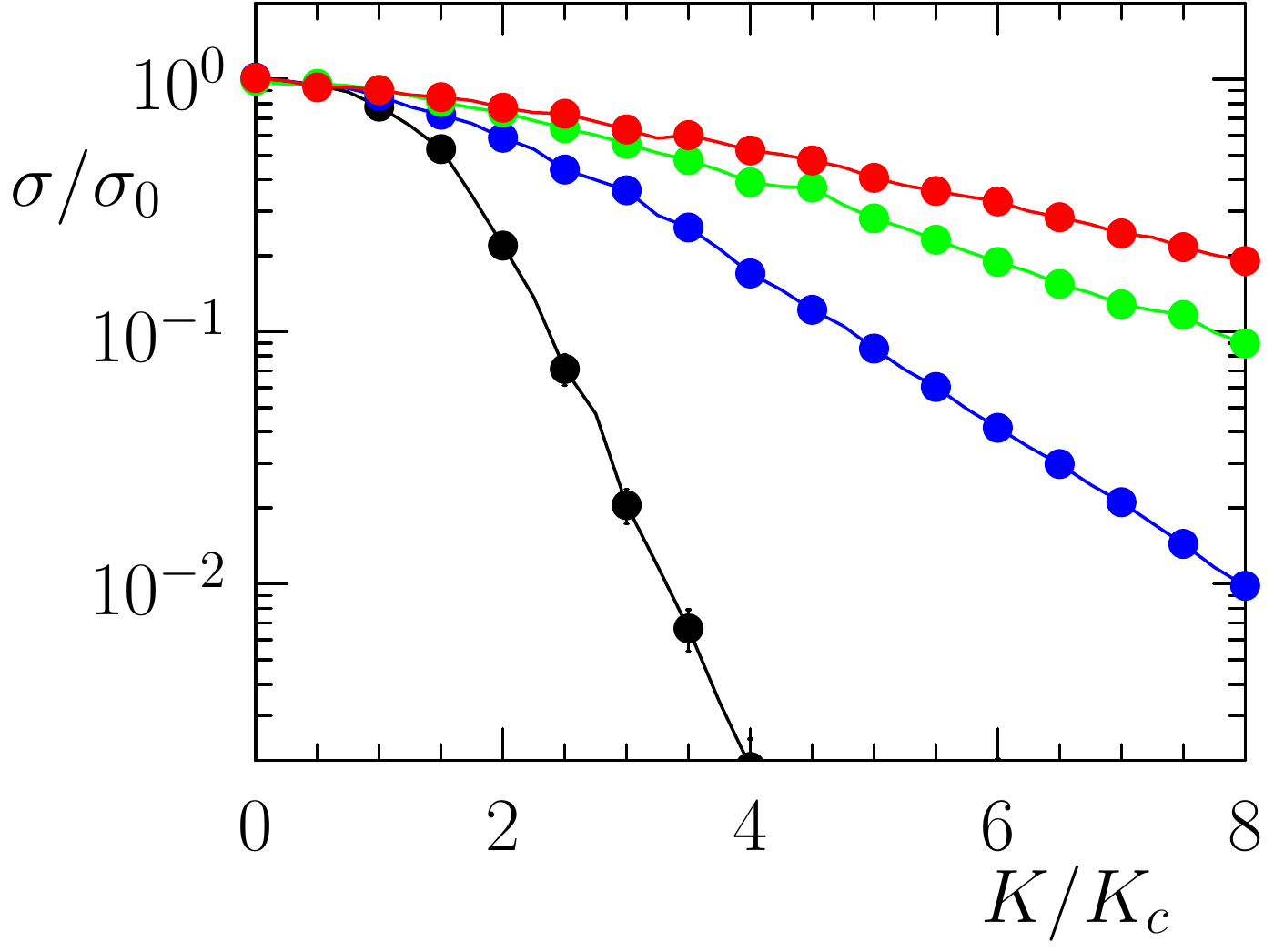}
	\end{center}
	\caption{\label{figS2}
		Dependence of the rescaled conductivity of the atom chain
		$\sigma/\sigma_0$ on the rescaled potential amplitude
		$K/K_c$ at different rescaled temperatures $T/K_c$. 
		Left panel: $T/K_c= 0.25$ (black curve),
		$1$ (blue curve), $2$ (green curve), $3$ (red curve),
		curves from bottom to top,
		$\nu = 34/21 \approx 1.618$, $K_c\simeq0.019$.
		Right panel: same values of $T/K_c$
		with the same order of curves, 
		$\nu = 55/21 \approx 2.618$, $K_c\simeq0.14$.
		Here $\sigma_0 =\nu /(2\pi \eta)$.
	}
\end{figure}

The dependence of the rescaled effective conductivity
$\sigma/\sigma_0$ is shown in Fig.~\ref{figS2}.

\begin{figure}[!htb]
	\begin{center}
		\includegraphics[width=0.49\columnwidth]{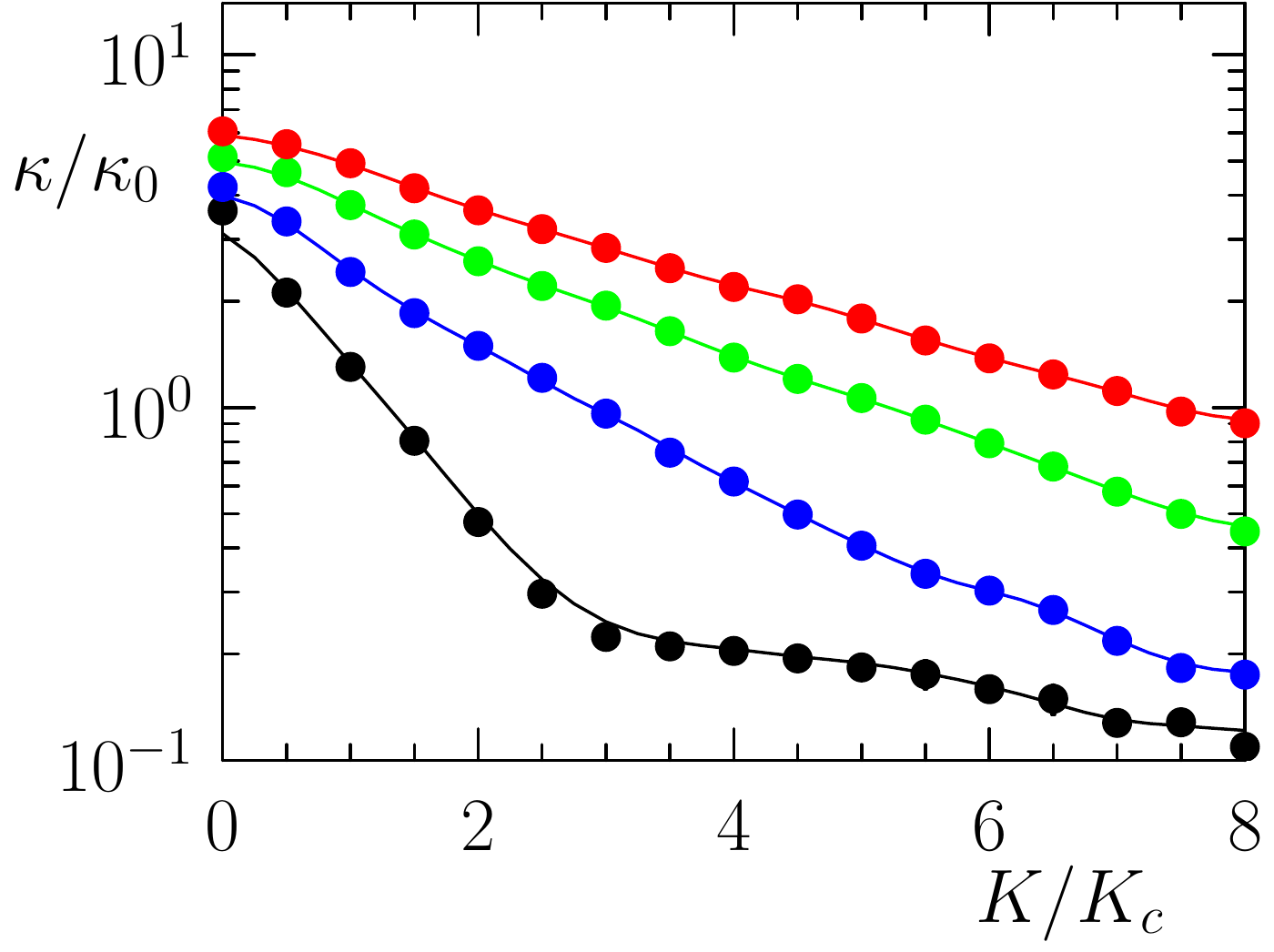}
		\includegraphics[width=0.49\columnwidth]{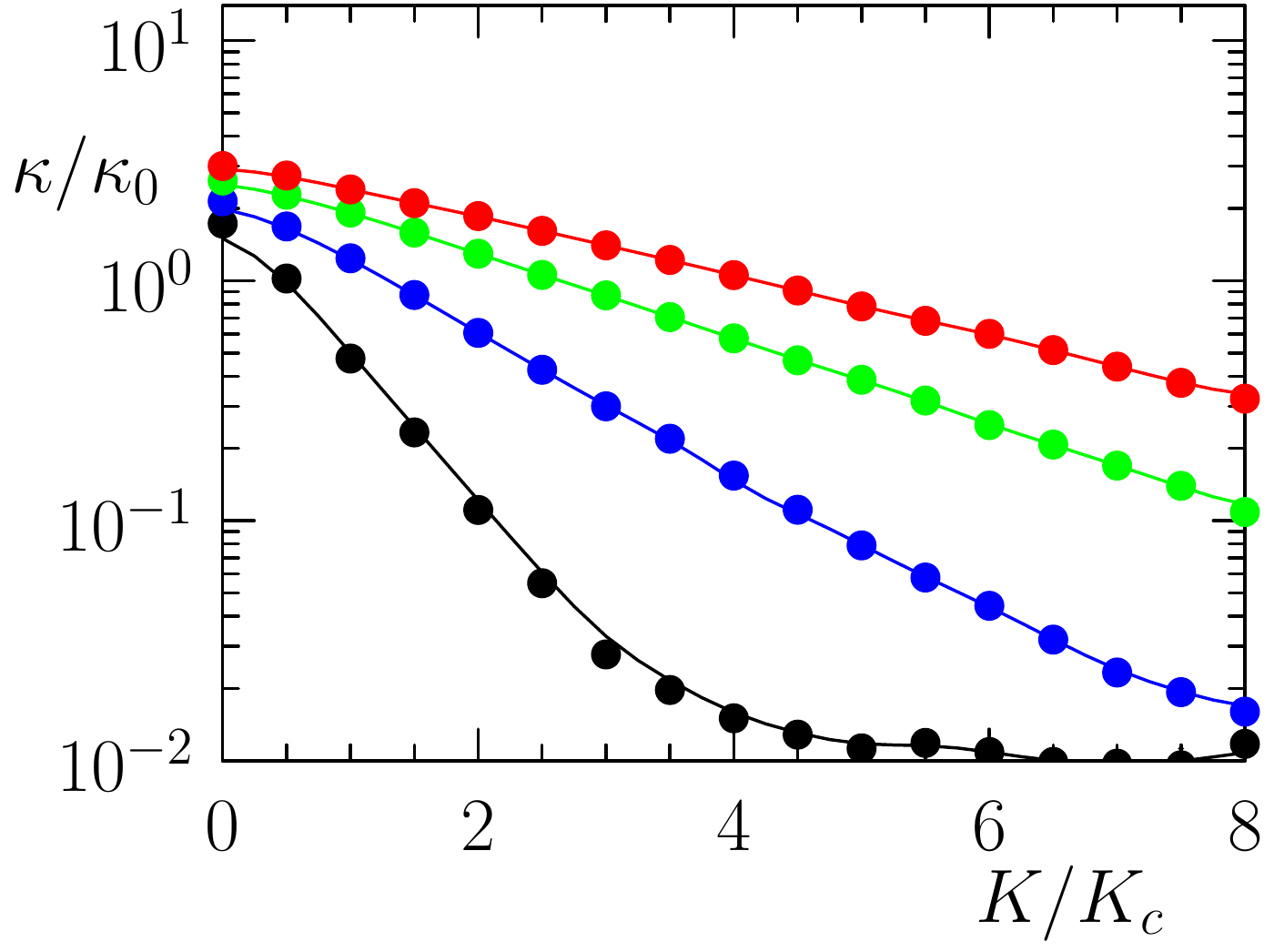}
	\end{center}
	\caption{\label{figS3}
		Dependence of the rescaled heat conductivity of the atom chain
		$\kappa/\kappa_0$ on the rescaled potential amplitude
		$K/K_c$ at different rescaled temperatures $T/K_c$. 
		The rescaled parameters are the same
		as in Fig.~\ref{figS2}.
		Here $\sigma_0 =\nu /(2\pi \eta)$,
		$\kappa_0=\sigma_0 K_c$.
	}
\end{figure}

The heat conductivity
is computed from the Fourier law
$J = \kappa(\partial T/ \partial x)$.
Here $J$ is the heat flow computed as described in
\cite{ztzs,lagesepjd}. Namely, it is computed from 
forces acting on a given atom $i$ 
from left and right sides being respectively
$f_i^{L}=\sum_{j<i} 3/|x_i-x_j|^4$,
$f_i^{R}=-\sum_{j>i} 3/|x_i-x_j|^4$.
For an atom moving with a velocity $v_i$,
these forces create  left and right energy flows 
$J_{L,R} = \langle f_i^{L,R} v_i \rangle_t \;$.
In a steady state, the mean atom energy is independent of 
time and $J_L + J_R=0$. But, the difference of these flows
gives the heat flow along the chain: 
$J=(J_R-J_L)/2 = \langle ( f_i^{R}- f_i^{L}) v_i/2 \rangle_t \; $.
Such  computations of the heat flow are done
with fixed atom positions at chain ends.
In addition, 
we perform time averaging
using accurate numerical integration along  atom trajectories
that cancels contribution of large oscillations 
due to quasi-periodic oscillations of atoms.
In this way, we  determine the thermal conductivity
via the relation $\kappa= 2\pi J L/\Delta T$.
The obtained results for $\kappa$ are independent
of small $\Delta T$. It is useful to compare
$\kappa$ with its value
$\kappa_0 = \sigma_0 K_c$.
The dependence of $\kappa/\kappa_0$ on $K$ at
different $T$
is shown in Fig.~\ref{figS3}.

\section{Independence of chain length}
\label{appendixD}

\begin{figure}[!htb]
	\begin{center}
		\includegraphics[width=0.49\columnwidth]{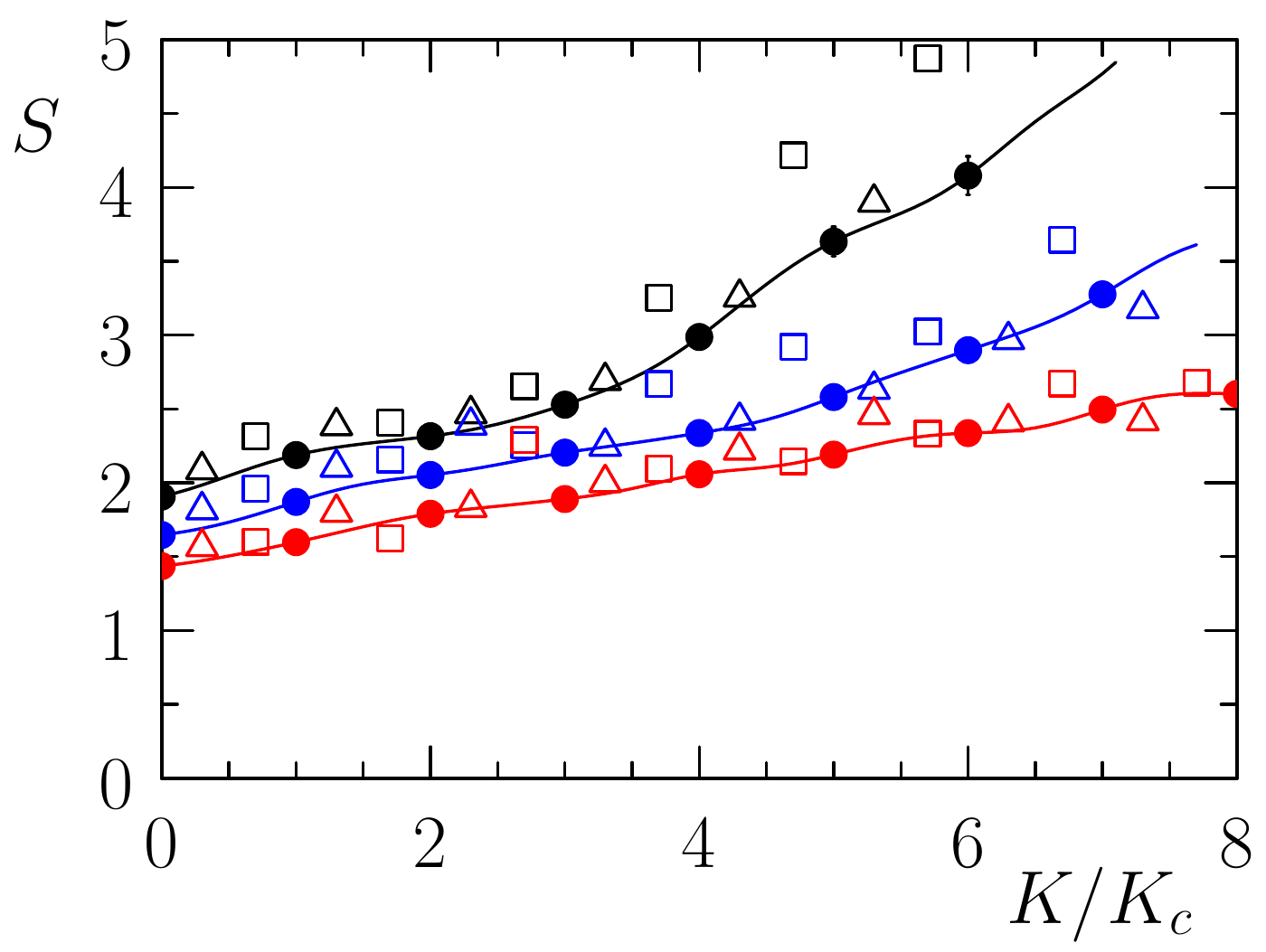}
		\includegraphics[width=0.49\columnwidth]{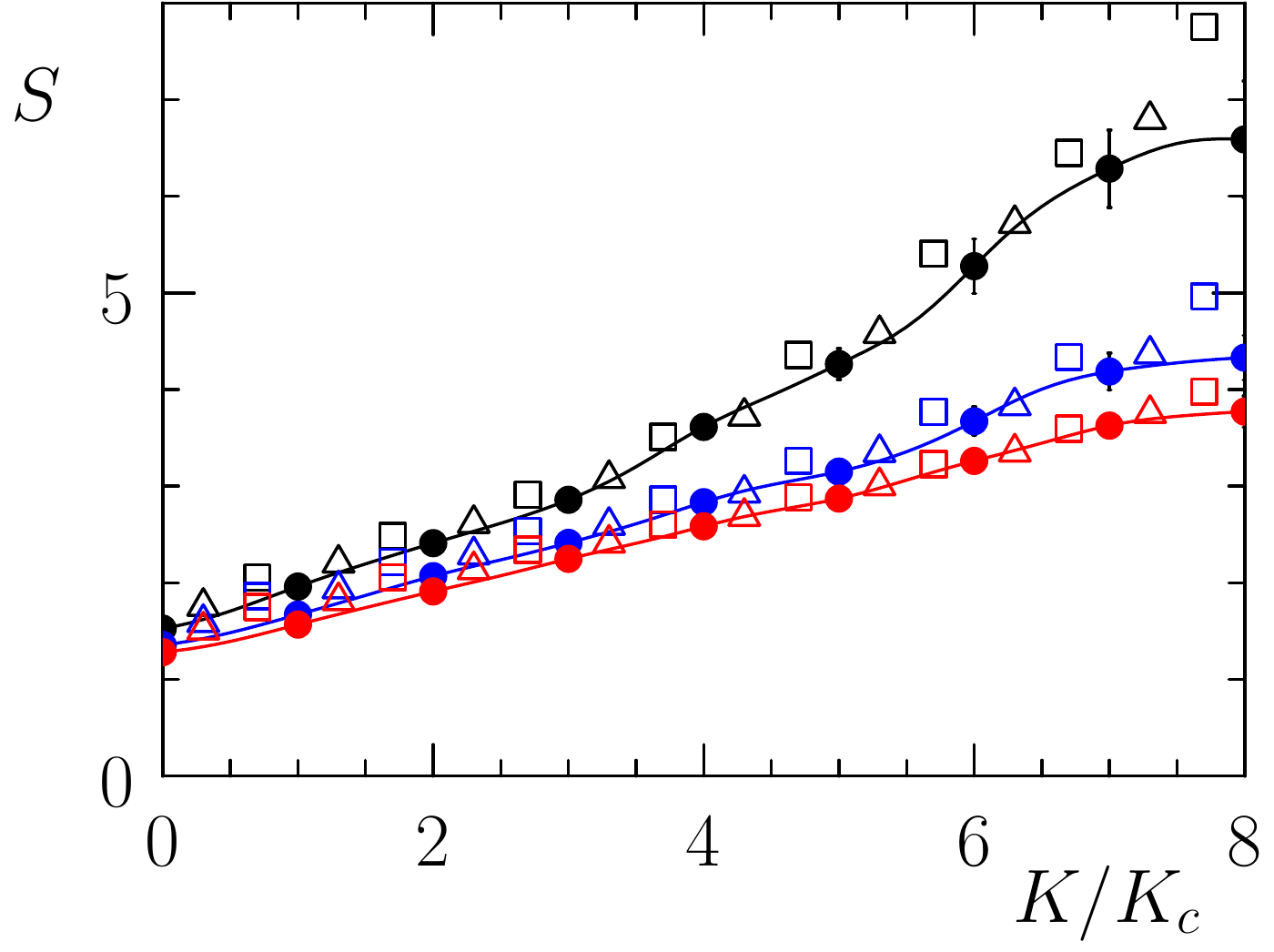}
	\end{center}
	\caption{\label{figS4}
		Dependence of the Seebeck coefficient $S$ 
		on the rescaled potential amplitude
		$K/K_c$ at different rescaled temperatures $T/K_c$. 
		Left panel: $T/K_c= 1$ (black curve/symbols),
		$2$ (blue curve/symbols), $4$ (red curve/symbols);
		here $\nu=N/L = 34/21 \approx 1.618$ (full circles),
		$55/34$ (triangles),
		$89/55$ (squares) with colors from bottom to top; 
		$K_c=0.019$.
		Right panel: same as in the left panel
		for $T/K_c = 0.8, 1.3, 1.6$
		at same color order
		for $\nu = N/L = 55/21, 89/34, 144/55 \approx 2.618$
		at same  symbol order;
		$K_c=0.14$.
	}
\end{figure}

Here in Fig.~\ref{figS4}, we present results at different chain lengths
for fixed atom density
showing that the Seebeck coefficient is independent of the system size.

We found similar independence of the chain length for $\sigma$, $\kappa$
and power factor $P_S=S^2 \sigma/\sigma_0$.

\section{Additional data for $ZT$}
\label{appendixE}

\begin{figure}[!h]
	\begin{center}
		\includegraphics[width=0.3\columnwidth]{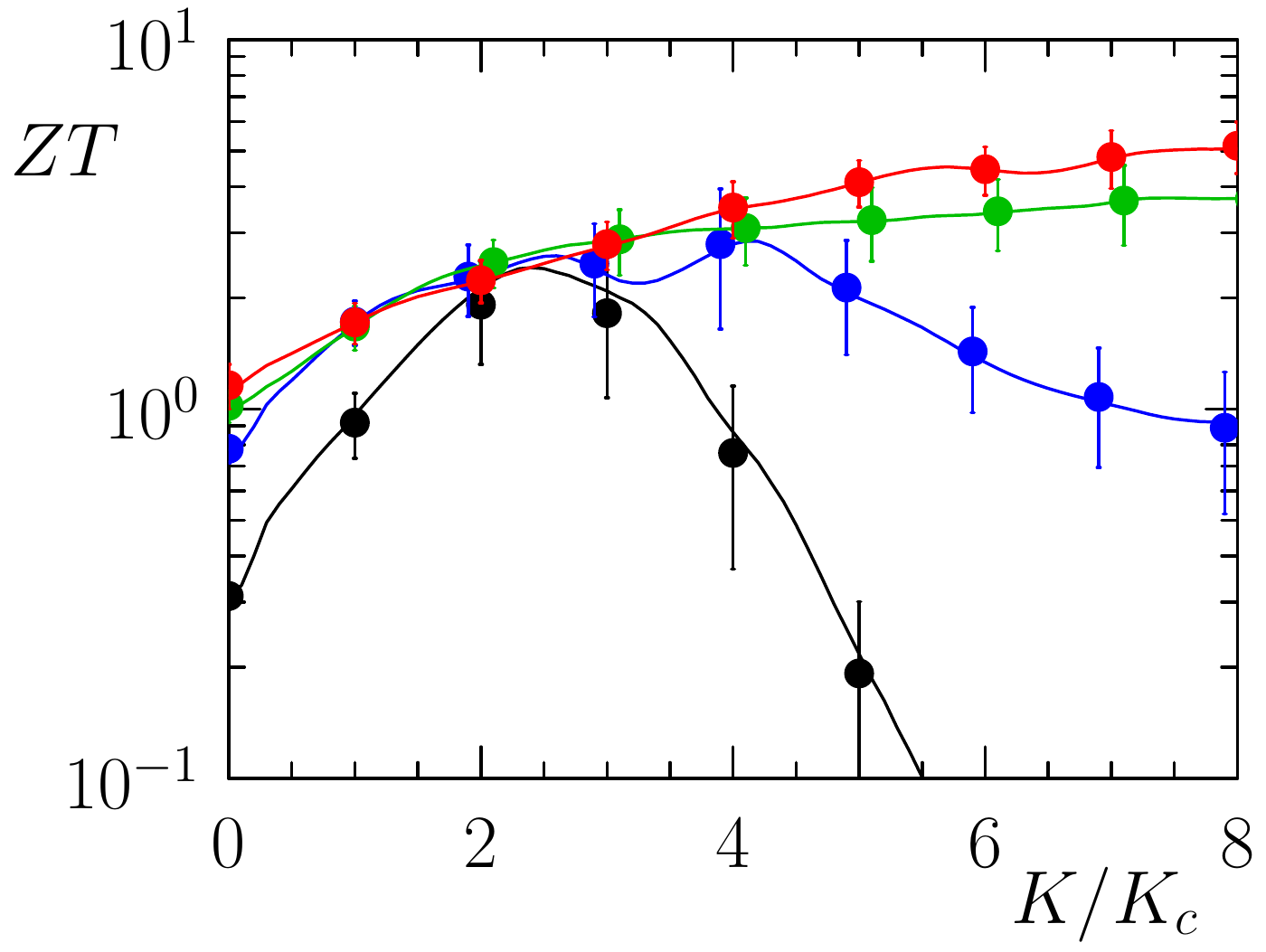}
		\includegraphics[width=0.3\columnwidth]{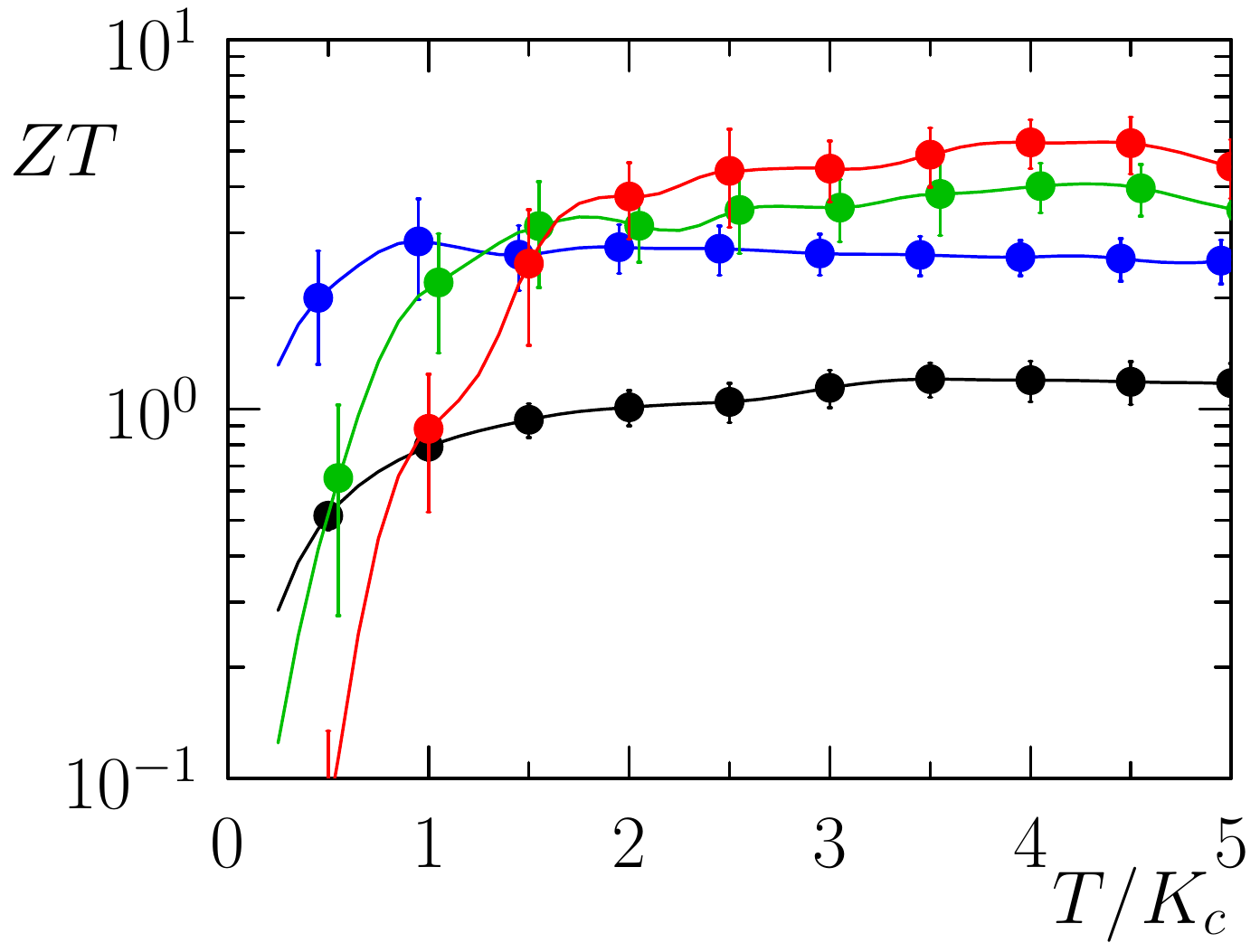}\\
		\includegraphics[width=0.3\columnwidth]{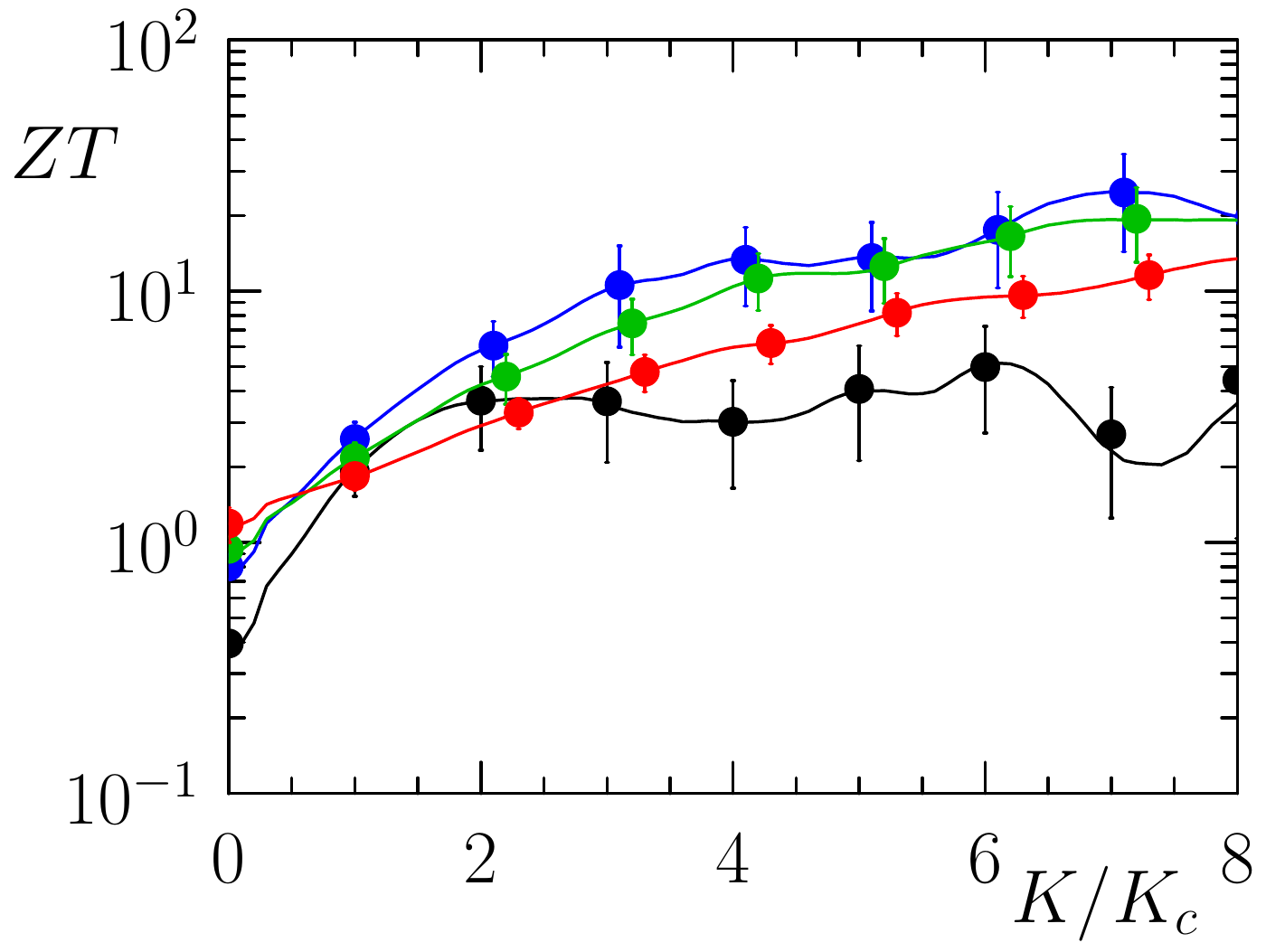}
		\includegraphics[width=0.3\columnwidth]{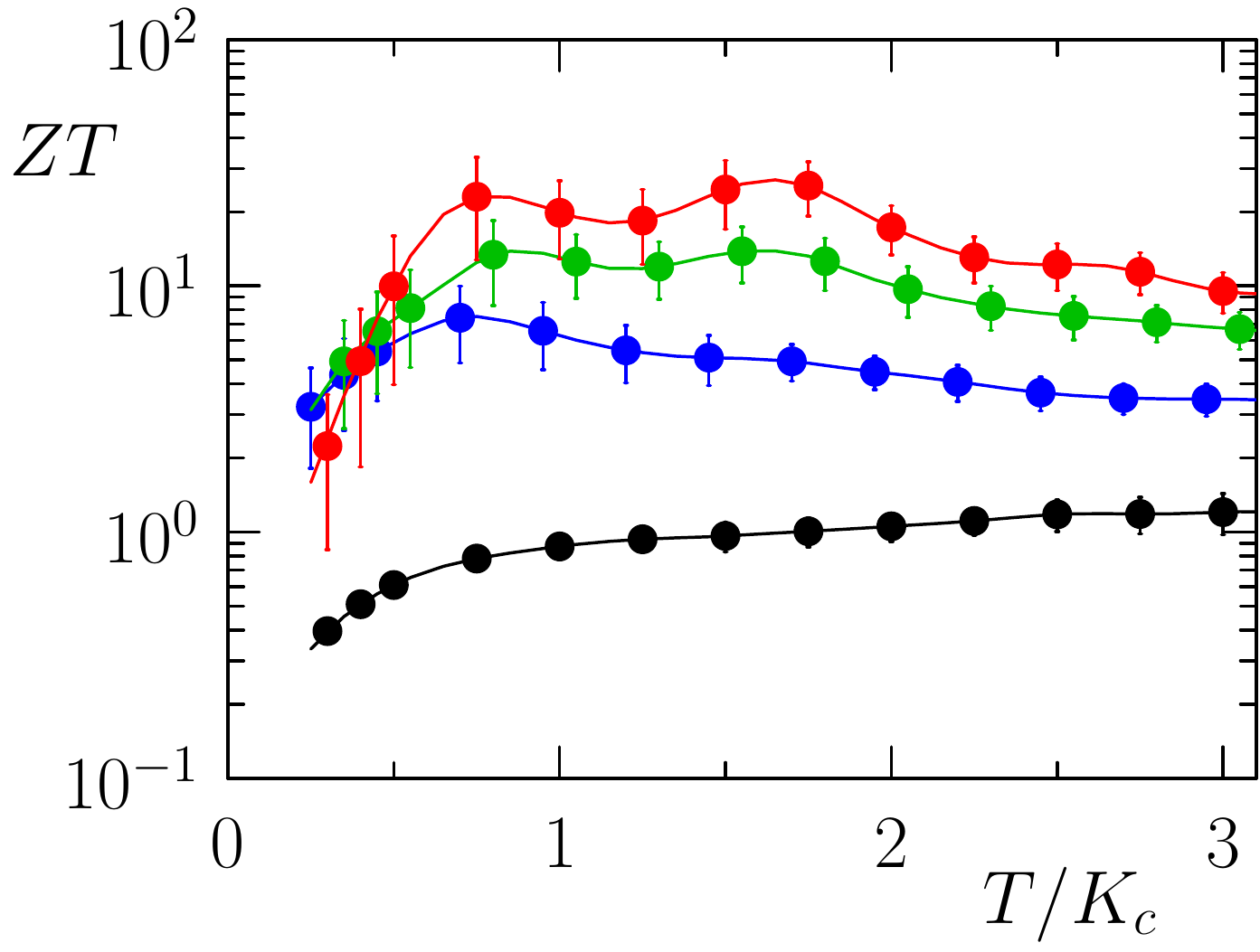}
	\end{center}
	\vglue -0.3cm
	\caption{\label{figS5}
		Dependence of  $ZT$ on the rescaled
		potential amplitude $K/K_c$ (left column)
		and on the rescaled temperature $T/K_c$ (right column)
		for $N/L=34/21 = \nu \approx 1.618$ (top row)
		and $N/L=55/21 = \nu \approx 2.618$ (bottom row).
		Left: color curves show the cases $T/K_c = 0.5, 1, 2, 3, 4$
		for $N/L=34/21$
		and $T/K_c = 0.3, 0.8, 1.3, 1.6, 2.3$
		for $N/L=55/21$
		(black, violet, blue, green, red from top to bottom curves).
		Right: $K/K_c= 0, 2, 4, 6, 8$ for $N/L=34/21$
		and  $N/L=55/21$
		(black, violet, blue, green, red from bottom to top curves).
		Here $K_c = 0.019$ for $\nu=34/21$ and $K_c=0.14$ for $\nu = 55/21$.
	}
\end{figure}

Additional data for the figure of merit $ZT$
are given in Fig.~\ref{figS5}.

\pagebreak
\section{Additional data for power factor}
\label{appendixF}

\begin{figure}[!ht]
	\begin{center}
		\includegraphics[width=0.49\columnwidth]{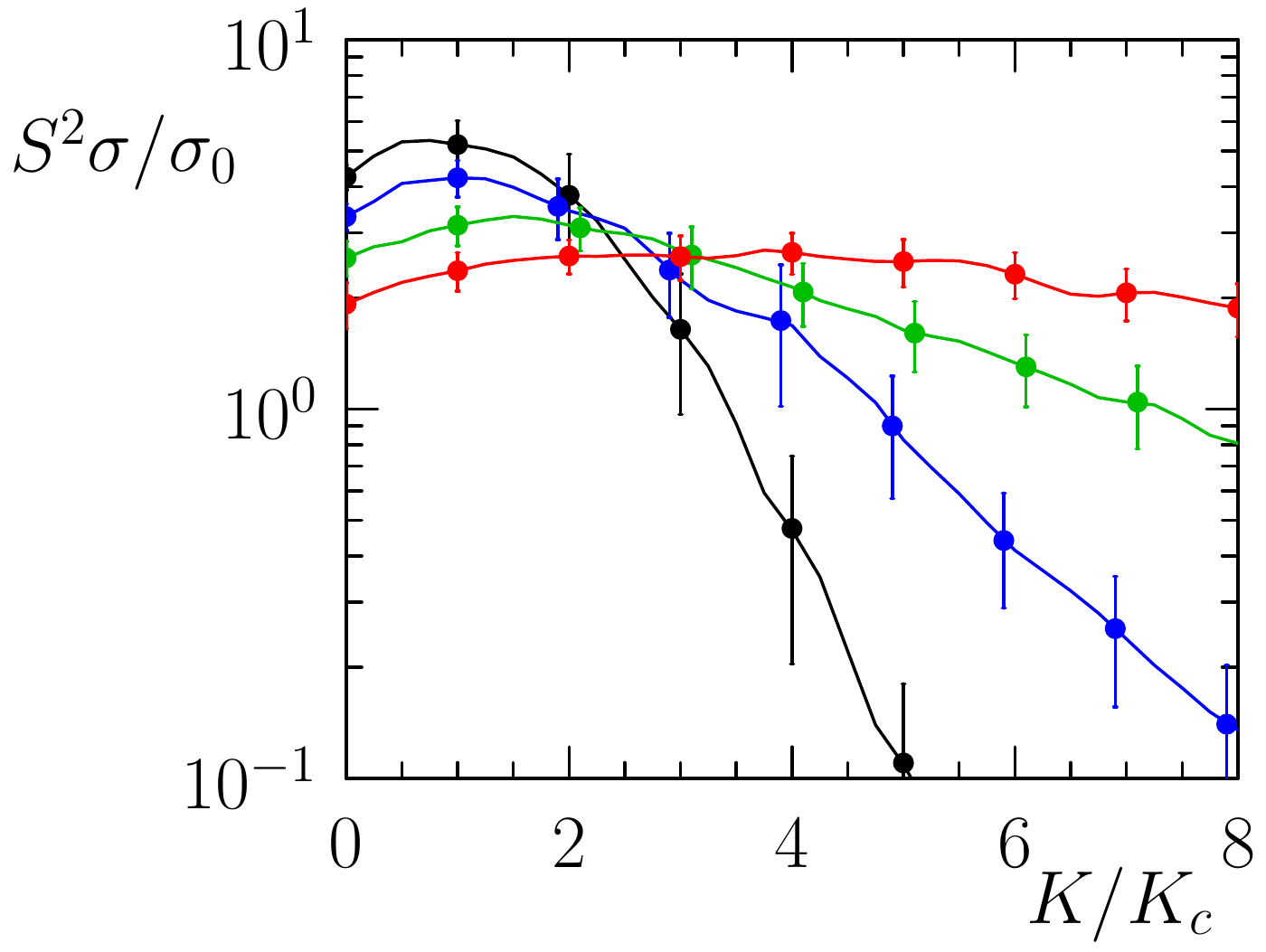}
		\includegraphics[width=0.49\columnwidth]{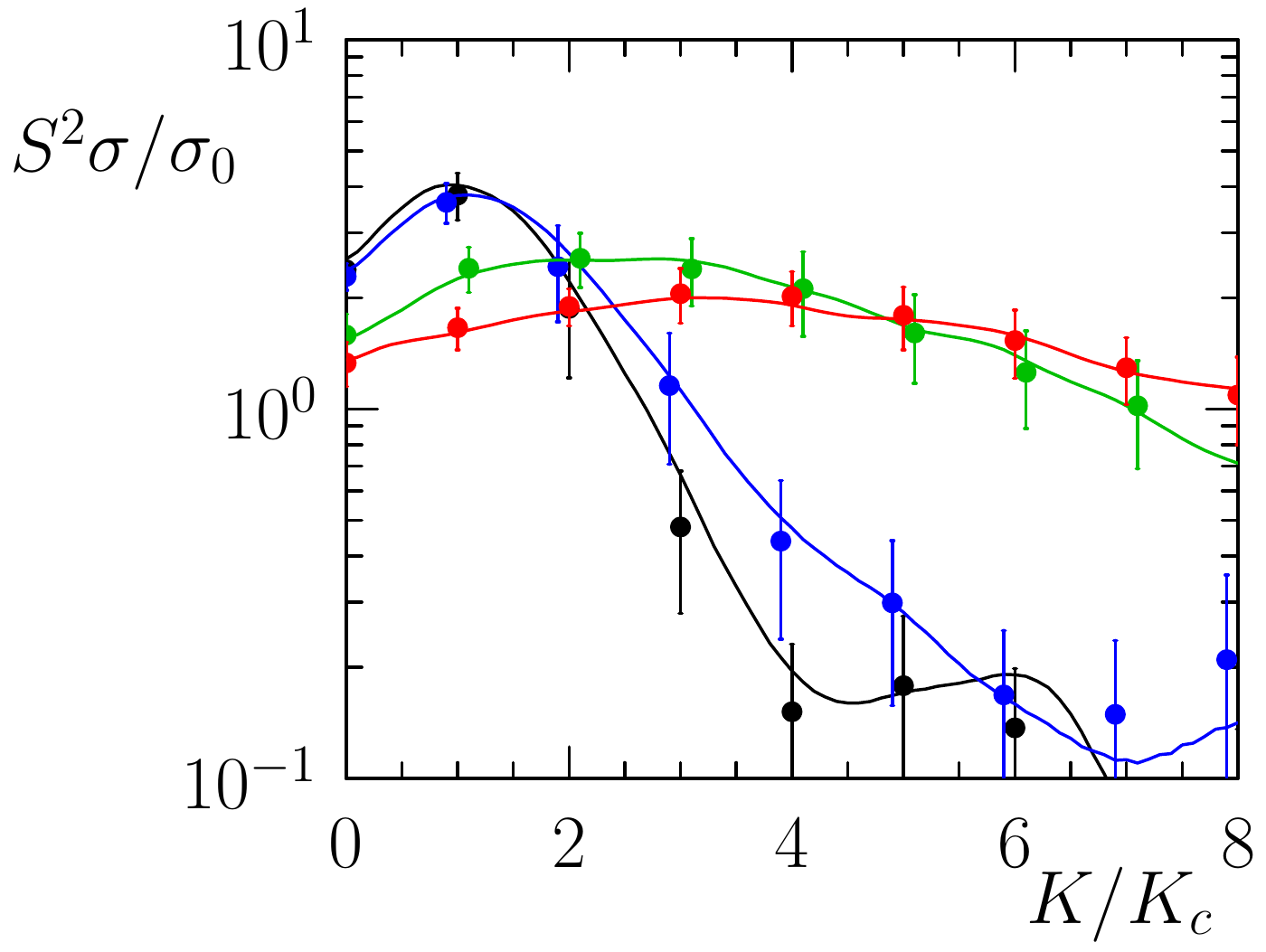}
	\end{center}
	\caption{\label{figS6}
		Dependence of the power factor $P_S=S^2 \sigma/\sigma_0$
		on system parameters.
		Left panel:
		$\nu = N/L = 34/21$, $T/K_c= 0.25, 1, 2, 4.25$
		(black, blue, green, red colors), $K_c=0.019$.
		Right panel:
		$\nu = N/L = 55/21$, $T/K_c= 0.25, 0.5, 1.5, 2.5$
		(black, blue, green, red colors), $K_c=0.14$.
	}
\end{figure}

Additional data for the power factor $P_S = S^2 \sigma/\sigma_0$
are given in Fig.~\ref{figS6}.

\end{document}